\documentclass    [%
reprint,
superscriptaddress,
 amsmath,amssymb,
 aps,
 prc,
]{revtex4-2}
\usepackage{graphicx}% Include figure files
\usepackage{dcolumn}% Align table columns on decimal point
\usepackage{bm}% bold math
\usepackage{subfigure}   %Lei20210525 multi-figures
\usepackage{color}   %Lei20210903 color mark, use \textcolor{red}{your txet}
\usepackage{ulem}   %Lei20211015 underline, strikeout, use \sout{text}
\usepackage{float}

\begin{document}

%\preprint{}

%\title{}

\title{ Vorticity and $ \Lambda $ polarization in the microscopic transport model PACIAE }   %Lei20210107


%\author{}
%\thanks{}
%\affiliation{}

%Lei20210107B
\author{ Anke Lei }
%\email[]{ ankeray@whut.edu.cn }
\affiliation{
    Department of Physics, Wuhan University of Technology, 
    Wuhan 430070, China.
}
\affiliation{
   Key Laboratory of Quark and Lepton Physics (MOE) and Institute of Particle Physics, Central China Normal University, Wuhan 430079, China.
}
\author{ Dujuan Wang }
\email{ wangdj@whut.edu.cn }
\affiliation{
    Department of Physics, Wuhan University of Technology, 
    Wuhan 430070, China.
}

\author{ Dai-Mei Zhou }
%\email[]{ zhoudm@mail.ccnu.edu.cn }
\affiliation{
   Key Laboratory of Quark and Lepton Physics (MOE) and Institute of Particle Physics, Central China Normal University, Wuhan 430079, China.
}
\author{ Ben-Hao Sa }
%\email[]{ sabh@ciae.ac.cn }
\affiliation{
   Key Laboratory of Quark and Lepton Physics (MOE) and Institute of Particle Physics, Central China Normal University, Wuhan 430079, China.
}
\affiliation{
   China Institute of Atomic Energy, P. O. Box 275 (10), Beijing, 102413 China.
}

\author{ Laszlo Pal Csernai }
%\email[]{ csernai@ift.uib.no }
\affiliation{
   Department of Physics and Technology, University of Bergen, Allegaten 55, 5007 Bergen, Norway.
}
\affiliation{
Frankfurt Institute for Advanced Studies, Ruth-Moufang-Strasse 1, 60438 Frankfurt am Main, Germany.
}
%Lei20210107E


%\date{\today}

%\begin{abstract}
%\end{abstract}

%Lei20210107B
\begin{abstract}
    We study the behavior of four types of vorticities in non-central Au+Au collisions at energies $ \sqrt{S_{NN}} $ = 5--200 GeV using a microscopic transport model PACIAE. The results show that the vorticities decay faster at lower energy and smaller impact parameter. The non-monotonic dependence of the initial vorticities on the collision energies is reconfirmed and the turning point is 10-15 GeV for different vorticities. The global $\Lambda$ polarization at energies $\sqrt{S_{NN}}$ = 7.7--200 GeV is reproduced by introducing an energy density freeze-out criterion.
\end{abstract}
%Lei20210107E

%\keywords{}

\maketitle

%\section{}
%\subsection{}
%\subsubsection{}

%Lei20210107B---
\section{\label{sec:intro} INTRODUCTION }
    The strongly coupled quark-gluon-plasma (sQGP) created in relativistic heavy-ion 
collisions produces a huge initial orbital angular momentum (OAM). It was proposed 
that such an OAM could be transformed to final hyperon polarization \cite{Liang:2004ph,Voloshin:2004ha}. Theoretical studies have been developed by statistical hydrodynamic approach based on the assumption that spin degrees of freedom are at local thermodynamic equilibrium \cite{Becattini:2007nd,Becattini:2007sr,Becattini:2013fla} and later recovered by Wigner function formalism in a quantum kinetic approach \cite{Fang:2016vpj}.
Thereby the polarization is connected to thermal vorticity and serves as a good probe. Hydrodynamic models have been widely used to investigate vorticity and polarization 
in heavy-ion collisions, such as PICR \cite{Csernai:2013bqa,Becattini:2013vja,Xie:2016fjj,Xie:2017upb,Xie:2021fjn}, ECHO-QGP \cite{Karpenko:2016jyx,Becattini:2021iol} and 3-FD model \cite{PhysRevC.100.014908,Ivanov:2020wak}. 
On the other hand, transport models, like APMT \cite{Jiang:2016woz,Lic:2017sl,Xia:2018tes,Shi:2017wpk,Wei:2018zfb} 
and UrQMD \cite{Karpenko:2017lyj,Vitiuk:2019rfv}, could provide microscopic properties of particle transport, which may extract more detailed information about vorticity and polarization. Recent reviews could be found in Refs.~\cite{Becattini:2020ngo,Karpenko:2021wdm,Huang:2020dtn}. The discovery of hyperon polarization provides a new way to understand Quark Matter (QM) in heavy-ion collisions.

    The measurement of hyperon polarization could be achieved through their parity violating weak decay in which the daughter particles emitted with a preferential direction \cite{Voloshin:2004ha}. STAR Collaboration has measured global and local $ \Lambda $ polarization in Au+Au collisions at the Relativistic 
Heavy-Ion Collider (RHIC) covering energy range $ \sqrt{S_{NN}} $ = 7.7--200 GeV \cite{STAR:2017ckg,Adam:2018ivw,Adam:2019srw,Adams:2021idn} and reported their 
dependence on collision energy, rapidity, transverse momentum and azimuthal angle. In Ref~\cite{Wu:2019eyi}, the choice of different types of vorticities gives the same azimuthal angle dependence of polarization as the experimental measurements, while the previous studies predicted the opposite trend \cite{Xia:2018tes,Becattini:2017gcx}. The observation of global $ \Lambda $ polarization shows evidence for the most vortical fluid produced in laboratory. 
ALICE Collaboration has measured global $ \Lambda $ polarization in Pb+Pb collisions at very high energy of $ \sqrt{S_{NN}} $ 
= 2.76 and 5.02 TeV \cite{Acharya:2019ryw}. Recently, the first measurement of global polarization of $ \Xi $ and $ \Omega $ 
hyperons in Au+Au collisions at $ \sqrt{ S_{NN} } $ = 200 GeV has been reported \cite{Adam:2020pti}. The new results give us possibility to compare 
polarizations for different particles and spins. It's worth noting that the HADES Collaboration reported a measurement of $ \Lambda $ polarization consistent with zero at a low energy $ \sqrt{S_{NN}} $ = 2.4 GeV \cite{Kornas:2020qzi}.
It may imply a non-monotonic trend in polarization versus increasing collision energy rather than the simply monotonic decline observed before \cite{STAR:2017ckg}. Such a trend may come from the non-monotonic dependence of the initial vorticities on the collision energy \cite{Deng:2020ygd}. Other studies predicted the maximum polarization would occur at  $\sqrt{S_{NN}} \approx$ 3 GeV \cite{Ivanov:2020udj} or 7.7 GeV \cite{Guo:2021uqc}. However, in two recent RHIC-STAR measurements, the turning point of global $\Lambda$ polarization has not yet been determined in Au+Au fixed-target collisions at $ \sqrt{S_{NN}} $ = 3 \cite{STAR:2021beb} and 7.2 GeV \cite{Okubo:2021dbt}. Nevertheless, because of the different energy scales, there may be other effects, for example, different hadronization scenarios. It still needs further studies.

    In this paper, we present the behavior of four types of vorticities, including their dependence on energy, time, centralities and spatial distribution, resulting from the microscopic transport model PACIAE \cite{Sa:2011ye}. The initial vorticity turning point is also studied and the $\Lambda$ polarization at 7.7-200 GeV is calculated.

    In order to facilitate the presentation, we will use the natural units $c=\hbar=k_{B}=1$, 
the Minkowskian metric tensor $g_{\mu\nu}=diag(1,-1,-1,-1)$ and the Levi-Civita symbol $\epsilon^{0123}=1$.

    The paper is organized as follow: a brief introduction to vorticity and polarization is given in Sec.~\ref{sec:theory}. Then our model setup and numerical method are described in Sec.~\ref{sec:model}. In Sec.~\ref{sec:res_dis} 
we will present and discuss our numerical results of the vorticity and the $\Lambda$ polarization. Finally, we give a summary in Sec.~\ref{sec:summary}. 
%\subsection{}
%\subsubsection{}

\section{\label{sec:theory} VORTICITY and POLARIZATION }
    In classical hydrodynamics, the flow vorticity is defined as
    \begin{equation}
        \bm{\omega} = \frac{1}{2} \boldsymbol{\nabla} \times \bm{v} ,
        \label{non_vor_vec}
    \end{equation}
    where $\bm{v}$ is the flow velocity. It is usually called non-relativistic vorticity (referred to as the NR-vorticity) and reflects the local angular velocity of a fluid cell. 
Its tensorial form can be written as
    \begin{equation}
            \omega^{NR}_{i j} = \frac{1}{2}\left(\partial_{i} v_{j}-\partial_{j} v_{i}\right) ,
            \label{Eq:non_vor_tens}
    \end{equation}
    where $v_{i}(i=1,2,3)$ denote the components of flow velocity. The unit of NR-vorticity is ${\rm [fm^{-1}] }$.
    
    The relativistic generalization of non-relativistic vorticity can be simply defined. 
There are three common forms: kinematic vorticity, temperature vorticity and thermal vorticity. In the following discussion, we will contain aboe four types of vorticities.  

    A pure generalization is kinematic vorticity (referred to as the K-vorticity) defined as
    \begin{equation}\label{Eq:K_vor_tens}
    \omega_{\mu \nu}^{K}=-\frac{1}{2}\left(\partial_{\mu} u_{\nu}-\partial_{\nu} u_{\mu}\right) ,
    \end{equation}
    where ,$u^{\mu}=$ $\gamma(1,\bm{v})$ is the four-velocity and $\gamma=1 / \sqrt{1-\bm{v}^{2}}$ is the Lorentz factor. Similarly, the unit of K-vorticity is ${\rm [fm^{-1}] }$.

    The temperature vorticity, also called T-voriticity, is defined as
    \begin{equation}\label{Eq:temp_vor_tens}
        \omega_{\mu \nu}^{T}=-\frac{1}{2}\left[\partial_{\mu}\left(T u_{\nu}\right)-\partial_{\nu}\left(T u_{\mu}\right)\right] ,
    \end{equation}
    where  $T$ is the local temperature of the fluid cell. The unit of T-vorticity is ${\rm [fm^{-2}] }$.

    The thermal vorticity is defined as
    \begin{equation}\label{Eq:th_vor_tens}
        \omega_{\mu \nu}^{th} \equiv \varpi_{\mu\nu} = -\frac{1}{2}(\partial_{\mu}\beta_{\nu} - \partial_{\nu}\beta_{\mu} ) ,
    \end{equation}
    where $ \beta^{\mu} = u^{\mu}/T $ is the inverse temperature four-vector field. The thermal vorticity is a dimensionless quantity and plays an important role in vorticity-induced particle polarization. The theoretical aspects of the vorticity-induced particle polarization can be found in Refs.~\cite{Liang:2004ph,Voloshin:2004ha,Fang:2016vpj}. Recently, new theoretical studies show that part of the polarization comes from the thermal shear tensor \cite{Becattini:2021suc,Liu:2021uhn}. However, as an extension of the PACIAE model, in this work, we do not consider this part of the contribution. We will follow the forms and notations in Ref.~\cite{Fang:2016vpj,Lic:2017sl} hereafter.

    The ensemble average of the spin four-vector for spin-$ s $ fermions 
with mass $ m $ and four-momentum $ p $ at space-time point $ x $  reads
    \begin{equation}\label{Eq:spin_pol_orig}
    S^{\mu}(x,p) = -\frac{s(s+1)}{6m}(1-n_{F})\epsilon^{\mu\nu\rho\sigma}p_{\nu}\varpi_{\rho\sigma}(x),
    \end{equation}
    where $ n_F = 1/(e^{\beta \cdot p \pm \mu/T} + 1 ) $ is Fermi-Dirac distribution of particles $(-)$ and antiparticles $(+)$ 
with $ \mu $ being relevant chemical potential and $ p_{0} = \sqrt{p^{2}+m^{2}} $. For dilute spin-$ 1/2 $ $ \Lambda $ 
and $ \overline{\Lambda} $ at much lower temperature of freeze-out $ ( m_{\Lambda/\overline{\Lambda}} \gg T)$, 
($ 1-n_{F} $) can be neglected. Hence, we will reach
    \begin{equation}\label{Eq:spin_pol_appr}
        S^{\mu}(x,p) = -\frac{1}{8m}\epsilon^{\mu\nu\rho\sigma}p_{\nu}\varpi_{\rho\sigma}(x) .
    \end{equation}

    In experiment, the $ \Lambda $ ( we will not distinguish between $\Lambda$ and $\overline{\Lambda}$ in the following text, unless specifically mentioned  ) polarization is measured in its rest frame and we 
denote it as $ S^{\ast\mu} = (0,\bf{S^{\ast}}) $ here. It is related to $ S^{\mu} $ in the ceter-of-mass frame of collision 
by a Lorentz boost
    \begin{equation}\label{Eq:spin_pol_boost}
        \mathbf{ S^{\ast} } (x,p) = \mathbf{S} - \frac{\mathbf{p\cdot S}}{p_{0}(m + p_{0})} \mathbf{ p } .
    \end{equation}

    In order to get the average spin vector, we need to take average of $ S^{\ast} $ over all $ \Lambda $ prduced at 
the hadronization stage:
    \begin{equation}\label{Eq:spin_pol_ave}
        \langle \mathbf{S^{\ast}} \rangle = \frac{1}{N}\sum_{i=1}^{N}\mathbf{S^{\ast}}(x_{i},p_{i}) ,
    \end{equation}
    where $ N $ is the total number of $ \Lambda $ collected in all events. 

    Finally, the spin polarization in three-direction $ \mathbf{n} $ is 
    \begin{equation}\label{Eq:spin_pol_normal}
        P_{n} = \frac{1}{s}\langle \mathbf{S^{\ast}} \rangle \cdot \mathbf{n}
    \end{equation}
    including the normalization factor $ 1/s $. For the global $\Lambda$ polarization, it will be projected on the direction of global angular momentum:
    \begin{equation}\label{Eq:spin_pol_glob}
        P=2 \frac{ \langle \mathbf{S^{\ast}} \rangle \cdot \mathbf{{J}}}{|\mathbf{J}|}
    \end{equation}

\section{\label{sec:model} MODEL }
    \subsection{\label{sec:PACIAE} PACIAE model }
    A microscopic parton and hadron cascade model PACIAE \cite{Sa:2011ye} is employed in present paper to simulate Au+Au 
collisions at $ \sqrt{s_{NN}} $ = 5--200 GeV. The PACIAE model is based on PYTHIA 6.4 events genetator \cite{Sjostrand:2006za}. However, PACIAE model simulates the parton initialization, parton rescattering, hadronization and the hadron rescattering sequentially. Therefore, the evolution of the high-energy collisions is described in more detail.

    In the PACIAE model, the nucleons in a nucleus-nucleus collision are distributed randomly in accordance with the collision geometry, number of participant(spectator) nucleons from the Glauber model and Woods-Saxon distribution. A nucleus-nucleus collision is then decomposed into many sub-nuleon-nucleon ($ NN $) collisions. A binary $ NN $ collision will happen if their relative transverse distance $D$ satisfies $ D \leq \sqrt{\sigma_{NN}^{tot}/\pi} $. The collision time is calculated with the assumption of straight-line trajectories. All of such $ NN $ collision pairs compose a $ NN $ collision list.

    An $ NN $ collision pair with the least collision time is selected from the collision list and executed by PYTHIA 6.4 (PYEVNW subroutine) with the string hadronization turned-off temporarily. The strings are broken up, and the diquarks (anti-diquarks) are split up randomly. Then the $ NN $ collision list is updated. By repeating the above routine for all selected $ NN $ collision pairs until the updated $ NN $ collision list becomes empty, the initial partonic state for a nucleus-nucleus collision is constructed. Then a partonic rescattering procedure with LO-pQCD parton-parton cross section is implemented. After partonic rescatteeing, the string is recovered and the hadronization is executed with the LUND string fragmentation regime. Finally, the system proceeds into the hadronic rescattering stage where only two-body collisions are considered. This results in the final hadronic state of the collision system. More details could be found in Ref.~\cite{Sa:2011ye}. It has to pointed out here that the PACIAE model does not introduce any phase transition assumption. Thus, if there is phase transition observed, it must be the result of dynamical evolution.
    
\begin{figure}[htbp]
    \centering
    %\flushleft
    \subfigure{
        \begin{minipage}[p]{1\linewidth}
        \centering
        \includegraphics[width=0.85\textwidth]{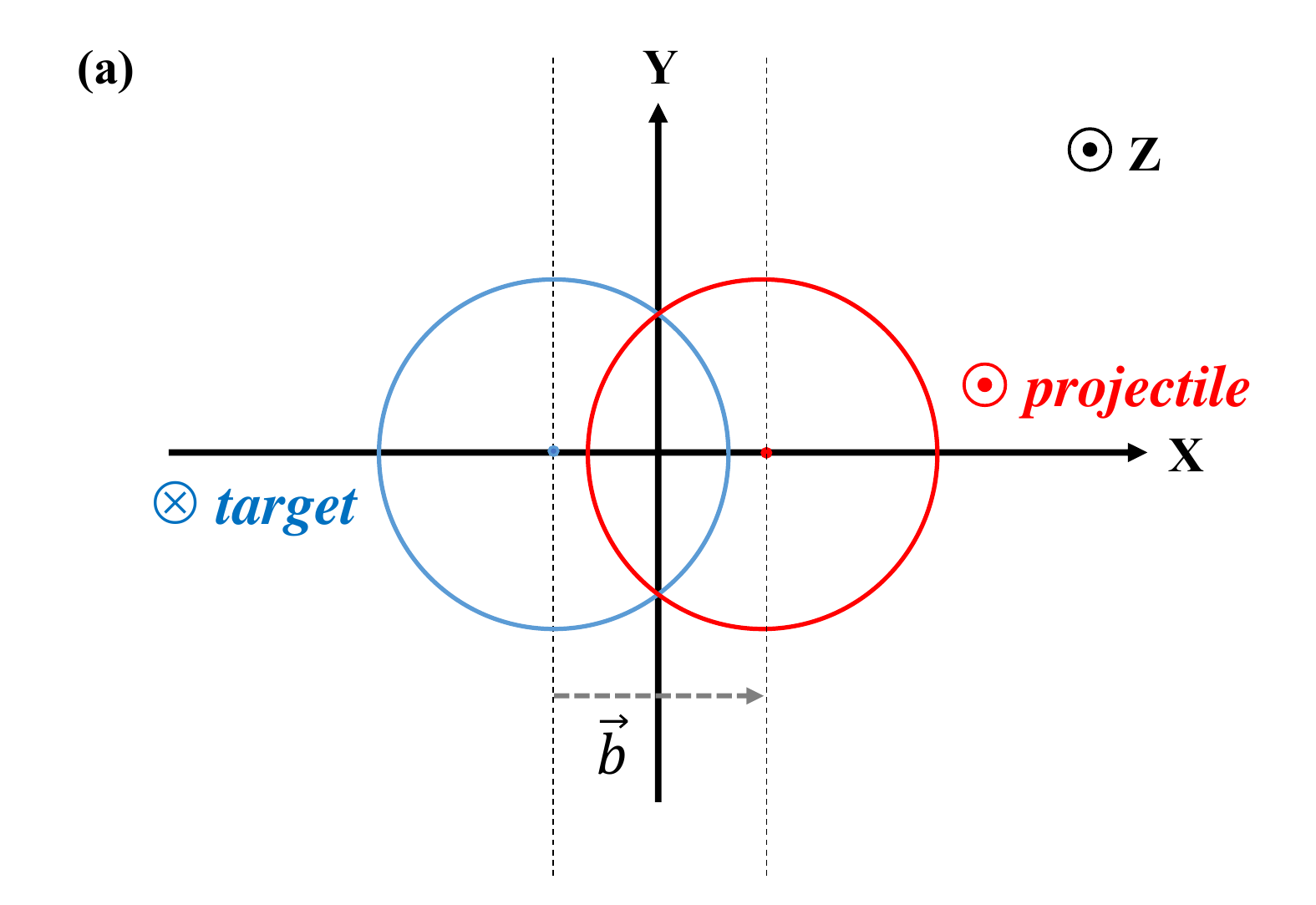}
        \includegraphics[width=0.85\textwidth]{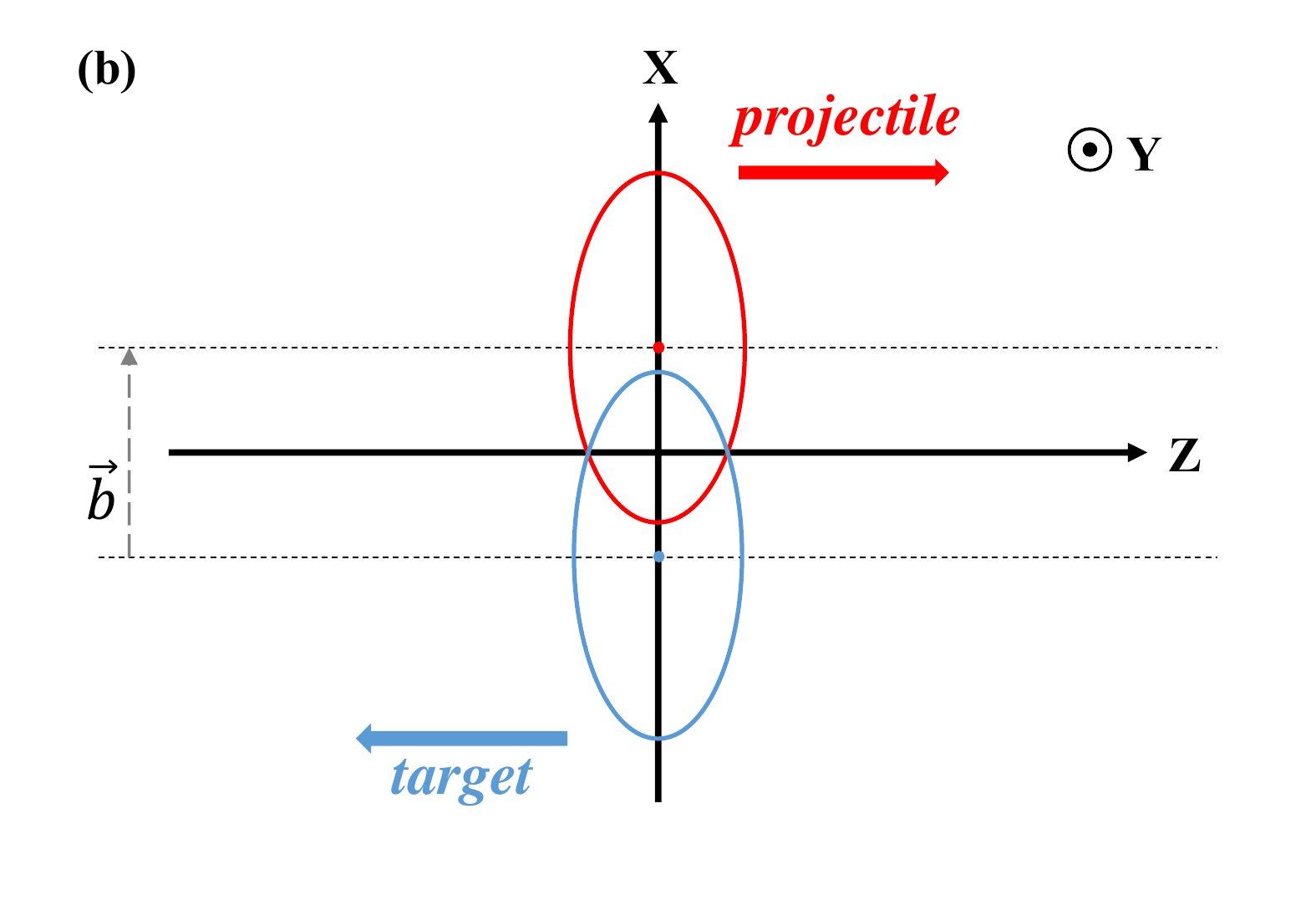}
        %\caption{}
        \end{minipage}
        }
    %    
    %\subfigure{
        %\begin{minipage}[p]{1\linewidth}
        %\centering
        %\includegraphics[scale=0.3]{Z_X_fig_color.pdf}
        %\caption{}
        %\end{minipage}
        %}
    \centering
    %\flushleft
    \caption{\label{coord_paciae}Schematic picture of the coordinate system in the PACIAE model. Panel (a) is in the transverse plane and (b) is in the reaction plane.}
    \end{figure}

    As Fig.~\ref{coord_paciae} shows, in the PACIAE model, the projectile (red circle) and target nucleus (blue circle) move along the +Z and -Z direction, respectively. The impact parameter vector $ \overrightarrow{b} $ points from the center of target nucleus to the center of projectile nucleus, along the X direction.
    \subsection{\label{sec:setup}  Model setup }
    The microscopic transport models such as AMPT, UrQMD and PACIAE allow us to track particles' positions and momenta. However, the flow velocity field cannot be obtained directly while it is natural in the hydrodynamic model. Thus, it is necessary to introduce a fluidized method. A simple and straightforward method is coarse-graining 
\cite{Teryaev:2015gxa,Jiang:2016woz,Lic:2017sl} and another widely applied one is introducing the smearing-function 
\cite{Deng:2016gyh,Pang:2012he,Hirano:2012kj,Oliinychenko:2015lva}. In the present paper, we employ a generalized coarse-graining method. We keep the default parameters values in the PACIAE model e.g. the two main external input parameters $K=1$ (the cross section correction factor) and $\beta=0.58$ (the parameter in LUND string fragmentation function). We calculate vorticities of the partonic stage and collect $ \Lambda$ from the hadronization stage to calculate the polarization.

    As Ref.~\cite{Teryaev:2015gxa,Jiang:2016woz,Lic:2017sl} described coarse-graining method, we firstly divide particles into grid-cells according their space-time coordinates. The cell size is set to be $ 0.5 \times 0.5 \times 0.5~\mathrm{fm^3} $, and the time slice is 0.5 $ \mathrm{fm/c}$. The energy density $ \epsilon $ and momentum density $ p $ of a cell will be obtained by summing over the corresponding quantities of all of the particles inside the cell.
To smooth too big fluctuation, more than $ 10^{4} $ events are generated for every condition we need, and take the events to average over those quantities.
Moreover, for structuring more continuous fluidization, we generalized the method as follows (taking a two-dimension case as an example, see Fig.~\ref{fig_Cg}): 
    \begin{figure}[htbp]
        \includegraphics[width=0.25\textwidth]{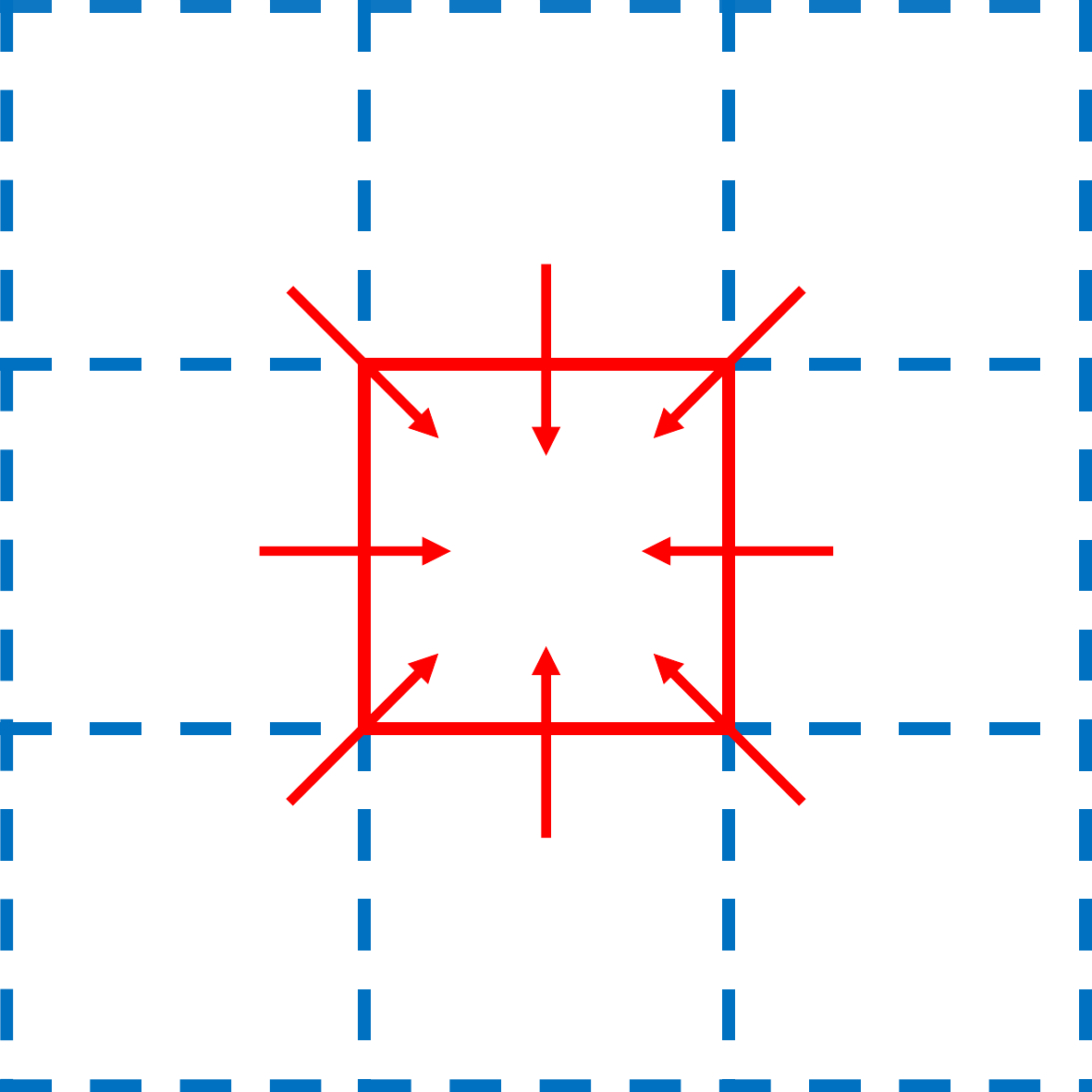}
        \caption{\label{fig_Cg} Schematic picture of the generalizaed coarse-graining.}
    \end{figure}
    
    We add $ \epsilon $ and $ p $ of each nearest four side cells and four corner cells into the central one and average the quantities over the number of these cells. Then we will get the corresponding average coarse-graining energy density and momentum density denoted as $ \overline{\epsilon} $ and $ \overline{p} $ respectively. A generalized coarse-graining fluidized map will eventually be obtained by repeating the procedure mentioned above to every cell of the system. (Cells with less than three particles will be discarded.) The flow velocity field is defined as $ \overline{p}/\overline{\epsilon} $. 
We note here that this generalized coarse-graining is totally different from increasing cell size simply, which only causes certain features of the fluid system to be obscured. On the other hand, the local cell temperature $T$
will be extracted from the local cell energy density $\overline{\epsilon}$ via the suitable relation \cite{Lin:2014tya}:
\begin{equation}\label{Eq:temp_parton}
    \overline{\epsilon} = \pi^{2}(16+10.5N_{f})T^{4}/30,
\end{equation}
where $ N_{f}=3 $ is number of considered quark flavors. 

\section{\label{sec:res_dis}  Numerical results }
    In this section, we will present our numerical results of the aforementioned four types of vorticities at $ \sqrt{S_{NN}} = $ 5--200 GeV and discuss the vorticities behaviors in the PACIAE model. In addition, the $\Lambda$ polarization in Au+Au collisions at  $ \sqrt{S_{NN}} = $  7.7--200 GeV is also presented. We focus on the vorticities along with the OAM, in the -Y direction. For characterizing the overall vorticity of the system, and eliminating errors on boundaries and low energy density areas, an energy-density-weighted average vorticity (referred to as the average vorticity) is used here: 
    \begin{equation}\label{Eq:ave_vor}
        \left< - \omega_{zx} \right>  \equiv  \left< - \omega_{Y} \right> = \frac{ \sum_{i}^{ N_{cell} } \overline{\epsilon_{i}} \omega_{i} }{ \sum_{i}^{N_{cell}} \overline{\epsilon_{i}} },
    \end{equation}
where $N_{cell}$ is the total number of the cells with non-zero energy density, $\overline{\epsilon_{i}}$ and $\omega_{i}$ are the energy density and the vorticity of the $i$-th cell. Different weighting methods will lead to different vorticity magnitude \cite{Jiang:2016woz,Csernai:2013bqa,Deng:2016gyh}, 
but it does not affect the relative trend characteristics of vorticities. 

\subsection{\label{subsec:dep_vorticity} The energy, time and centrality dependence of vorticity }
\begin{figure}[htbp]
    %\centering
    \flushleft
    %\flushright
    %\subfigure[NR-vorticity]{
    \subfigure{
        \begin{minipage}[t]{0.5\linewidth}
        %\centering
        %\flushleft
        \includegraphics[scale=0.2]{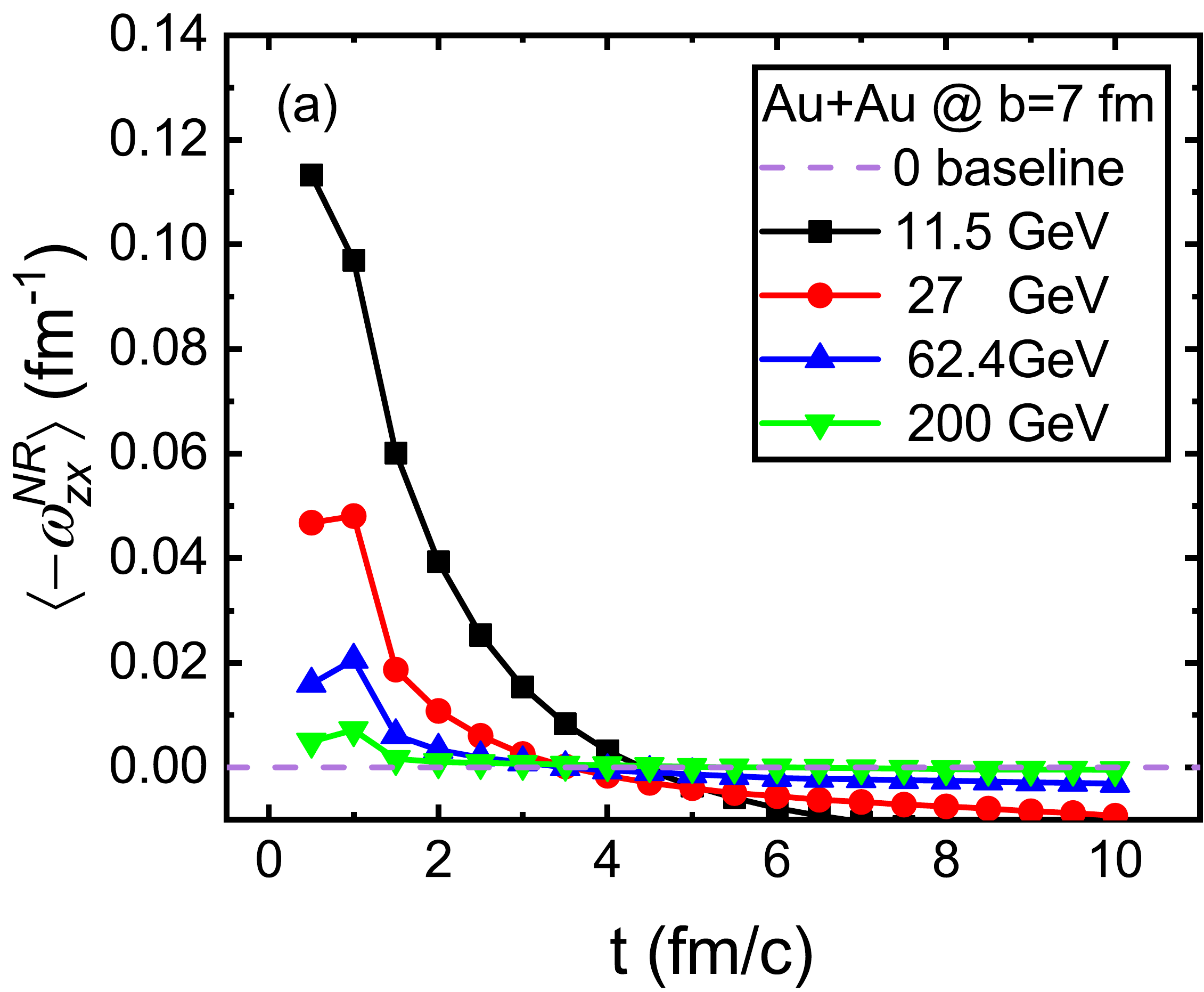}
        %\caption{NR_vor}
        %\includegraphics[scale=0.2]{kVorVsTime.pdf}
        %\caption{K_vor}
        %\includegraphics[scale=0.3]{tVorVsTime.pdf}
        %\caption{T_vor}
        %\includegraphics[scale=0.2]{thVorVsTime.pdf}
        %\caption{th_vor}
        \end{minipage}
        }
    %\subfigure[K-vorticity]{
    %\vfill
    \subfigure{
        \begin{minipage}[t]{0\linewidth}
        %\centering
        %\flushright
        \includegraphics[scale=0.2]{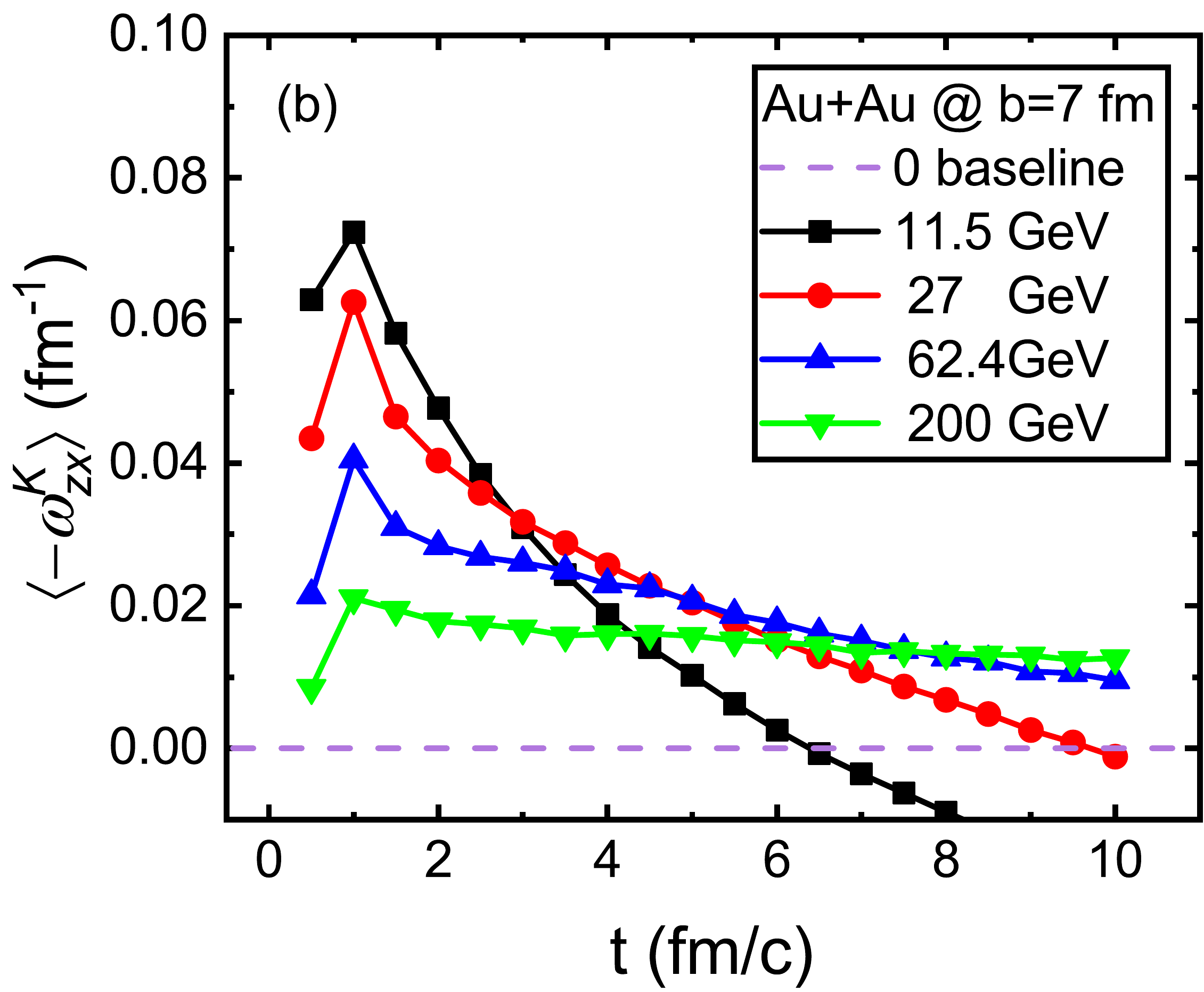}
        %\caption{K_vor}
        %\includegraphics[scale=0.3]{thVorVsTime.pdf}
        %\caption{th_vor}
        \end{minipage}
        } \\
        %} 
    %\vfill
    %\subfigure[T-vorticity]{
    \subfigure{
        \begin{minipage}[t]{0.5\linewidth}
        %\centering
        %\flushleft
        \includegraphics[scale=0.2]{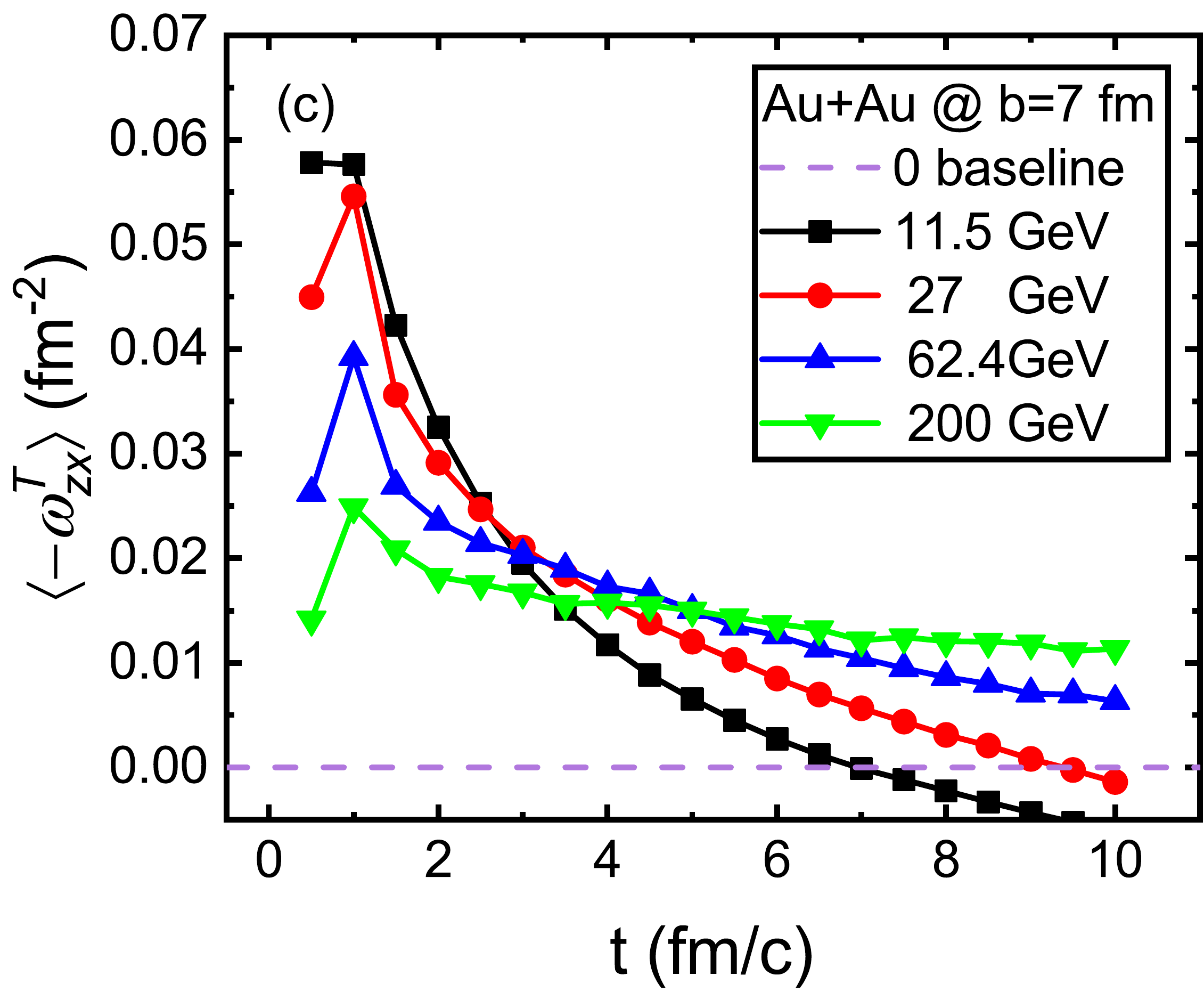}
        %\caption{T_vor}
        \end{minipage}
        }
    %\vfill
    %\subfigure[th-vorticity]{
    \subfigure{
        \begin{minipage}[t]{0\linewidth}
        %\centering
        %\flushright
        \includegraphics[scale=0.2]{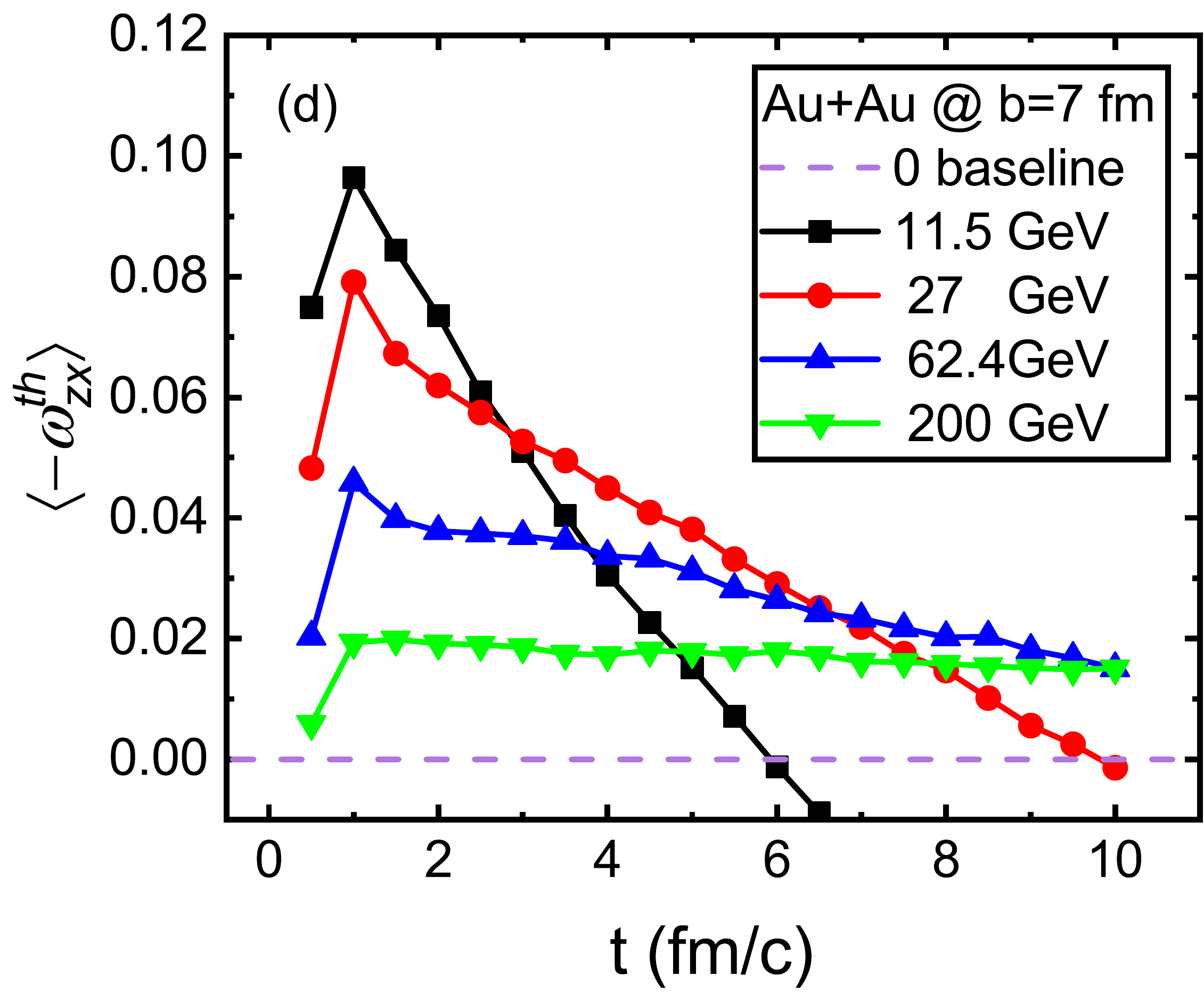}
        %\caption{th_vor}
        \end{minipage}
        }
    %\centering
    \flushleft
    %\flushright
    \caption{\label{fig:vor_vs_cmsE_time} Time evolution of the four types of average vorticities at different energies in the PACIAE model.}
    %\end{figure}
    \end{figure}

    We ignore the partonic initialization time so that our time origin begins from the partonic rescattering stage where QGP starts to form and evolve.
Fig.~\ref{fig:vor_vs_cmsE_time} shows the time evolution of four types of vorticities at different collision energies with impact parameter b=7 fm. One can see similar behaviors among four types of vorticities: 

(1) The magnitude is almost reached the peak value at 1 fm/c then decays with time. The behavior is consistent with other studies about K-vorticity and thermal vorticity \cite{Ivanov:2020wak,Jiang:2016woz,Deng:2016gyh,Wei:2018zfb}  which is attributed to the explosion of the QGP system. 

(2) The magnitude is smaller at higher energy before some points of time. Especially the NR-vorticity behaves more regularly than other types of vorticities. However, the NR-vorticity, without considering the relativistic correction, is not in line with the actual situation. In the
\begin{figure}[htbp]
    %\centering
    \flushleft
    %\flushright
    %\subfigure[NR-vorticity]{
    \subfigure{
        \begin{minipage}[t]{0.5\linewidth}
        %\centering
        %\flushleft
        \includegraphics[scale=0.2]{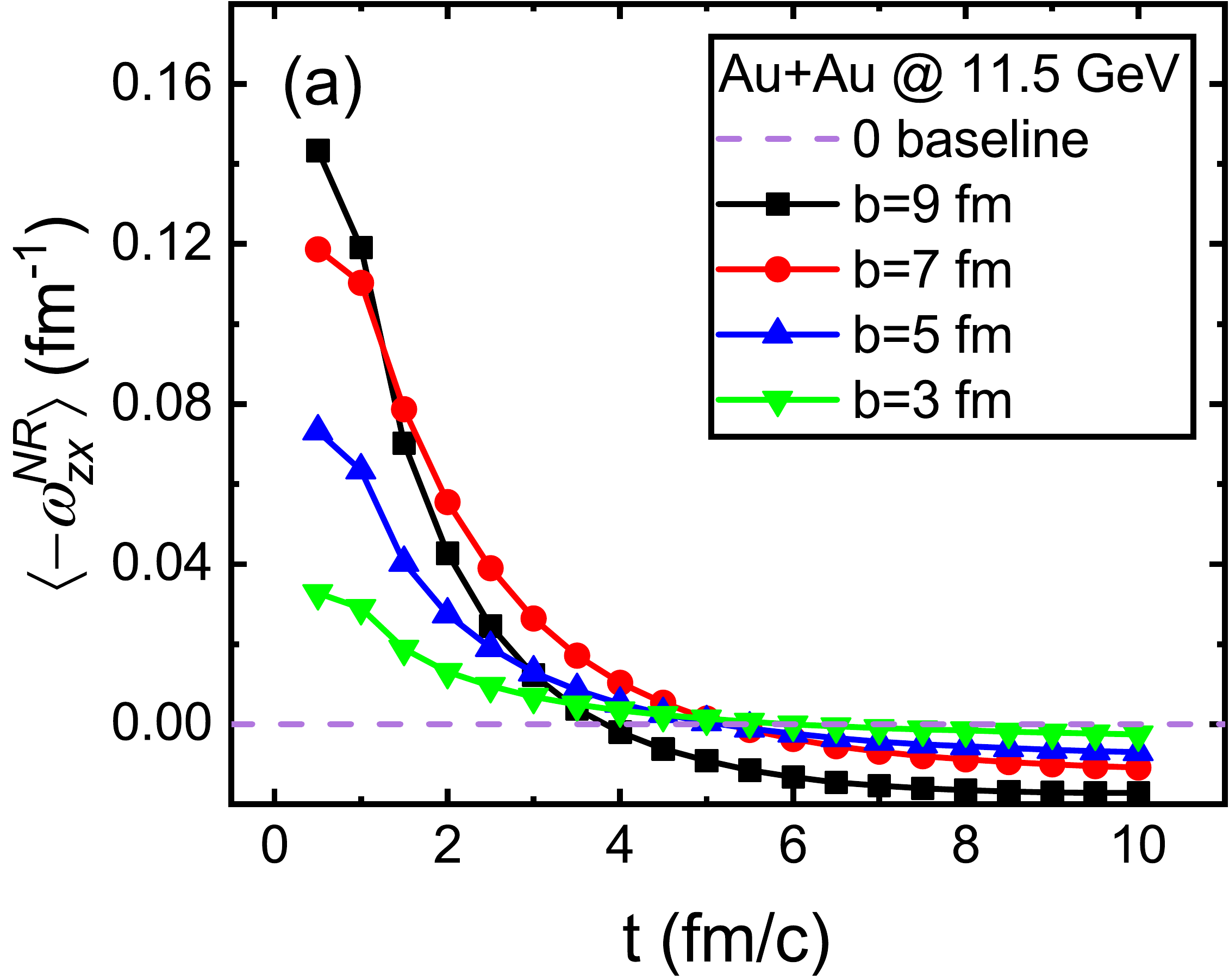}
        %\caption{NR_vor}
        \end{minipage}
        }
    %\subfigure[K-vorticity]{
    \subfigure{
        \begin{minipage}[t]{0\linewidth}
        %\centering
        %\flushright
        \includegraphics[scale=0.2]{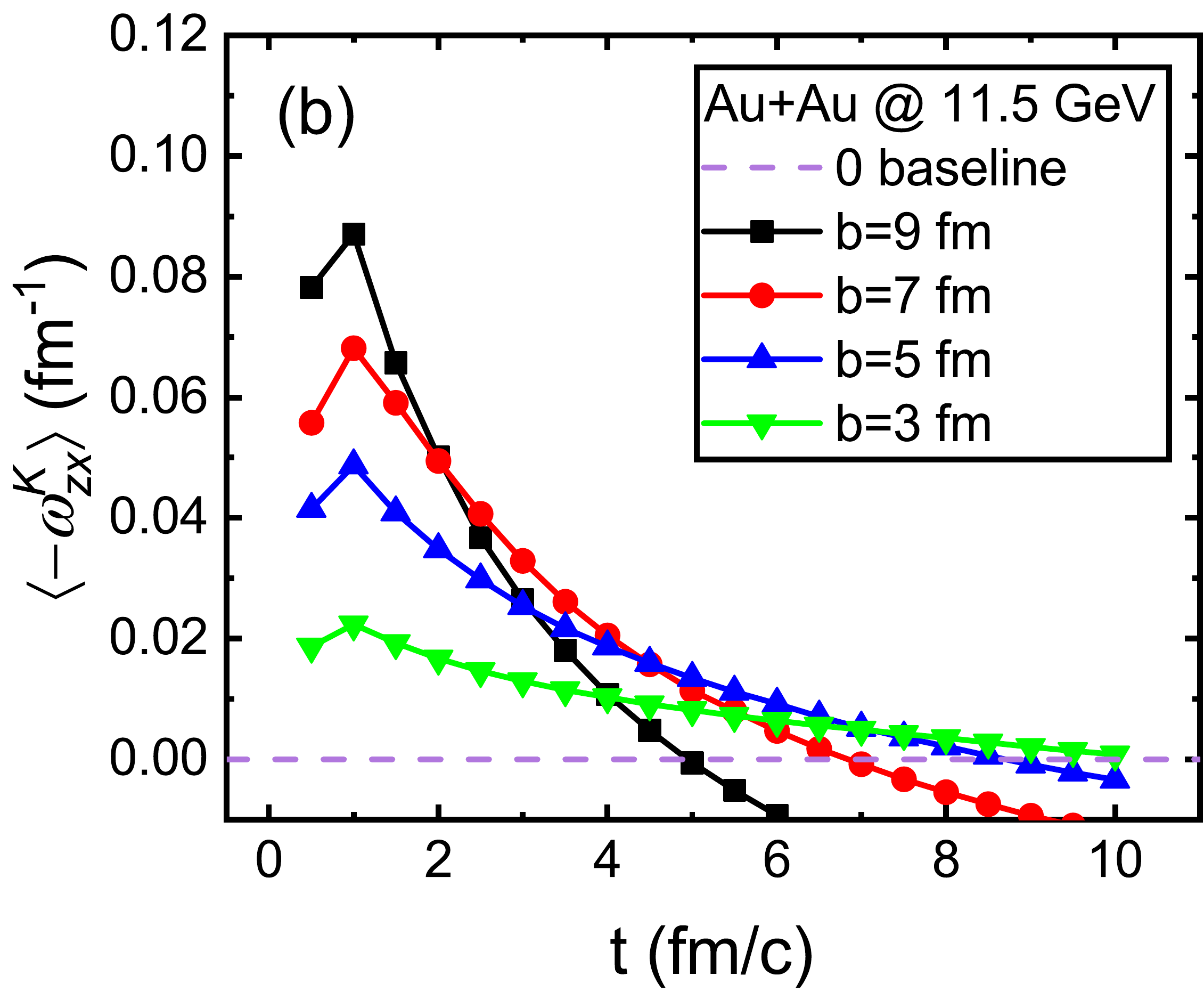}
        %\caption{K_vor}
        \end{minipage}
        } %\\
    \hfill
    %\subfigure[T-vorticity]{
    \subfigure{
        \begin{minipage}[t]{0.5\linewidth}
        %\centering
        %\flushleft
        \includegraphics[scale=0.2]{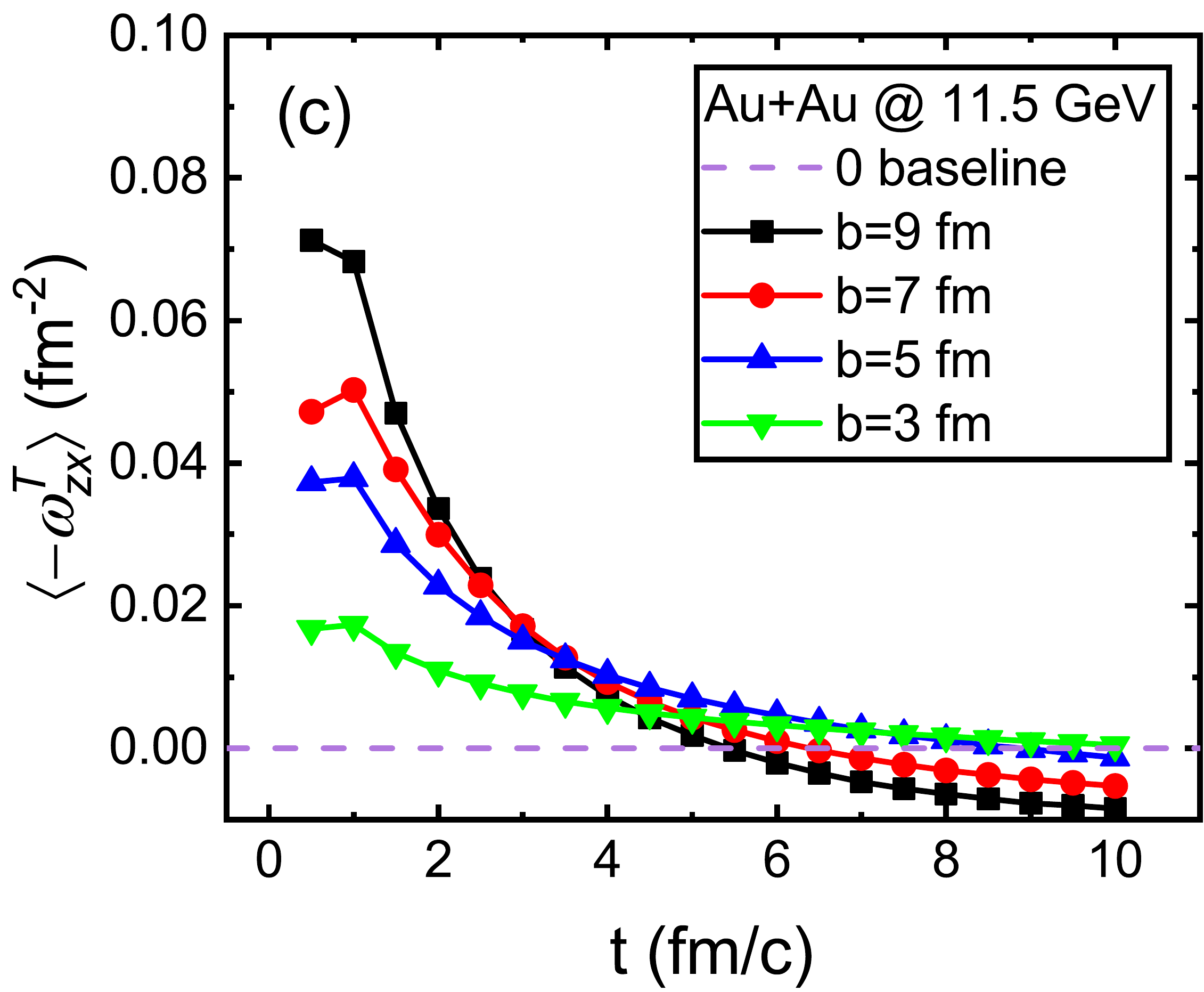}
        %\caption{T_vor}
        \end{minipage}
        }
    %\subfigure[th-vorticity]{
    \subfigure{
        \begin{minipage}[t]{0\linewidth}
        %\centering
        %\flushright
        \includegraphics[scale=0.2]{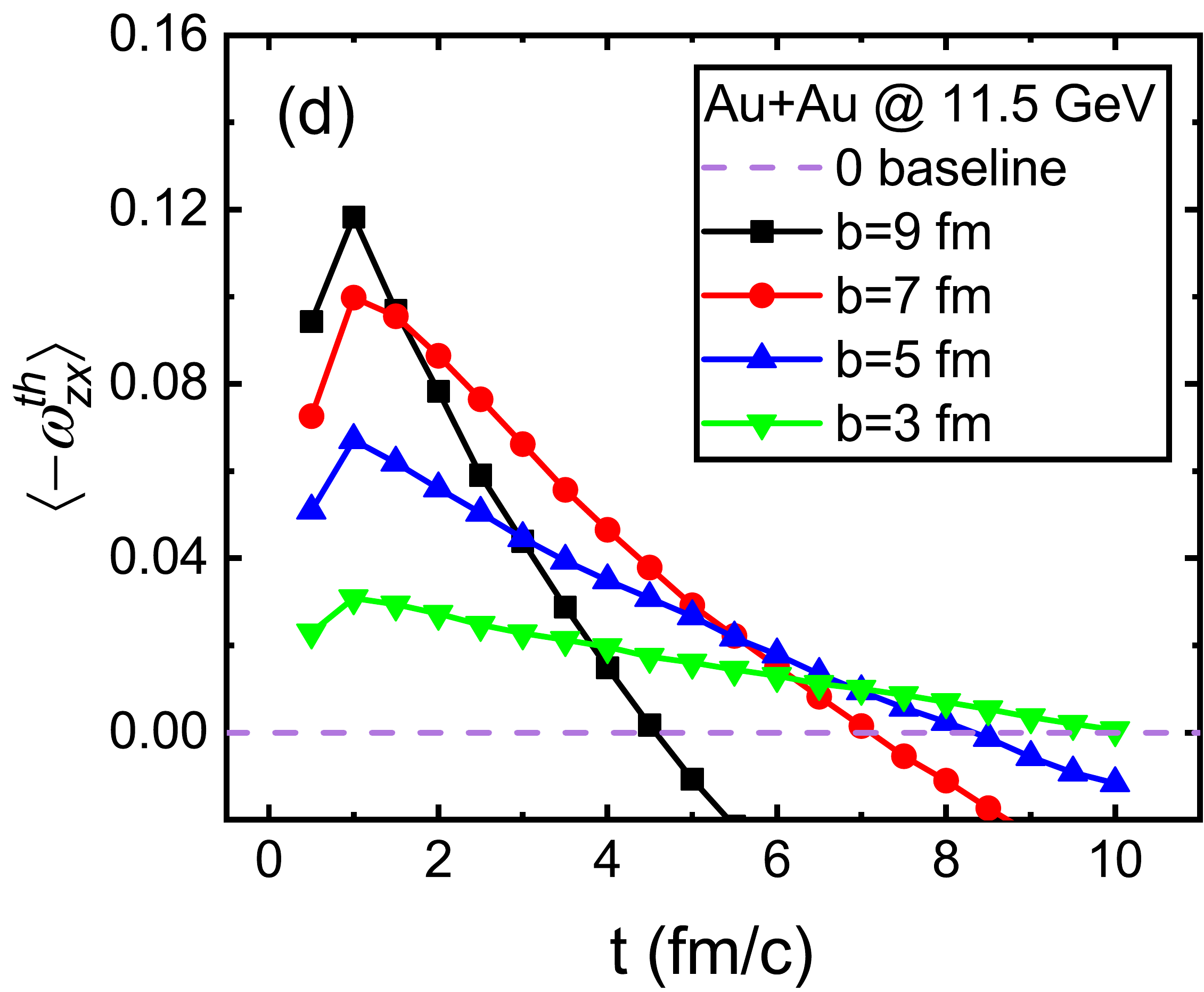}
        %\caption{th_vor}
        \end{minipage}
        }
    %\centering
    \flushleft
    %\flushright
    \caption{\label{fig:vor_vs_bPara} Time evolution of the four types of average vorticities at different impact parameters in the PACIAE model.}
    \end{figure}relativistic case, the other three vorticities show a faster decay trend at lower energy. This can be understood from the fact that the collision is less violent at low energy comparing with the higher energy one, and the formed QGP has a shorter time to exist and evolve.

(3)There are ``zero vorticities points'' that the average vorticities become zero even a reverse sign after some time due to the very low energy density weight. Such ``zero vorticities points'' appear earlier at lower energy. This also shows that the lower energy QGP system has a shorter evolution.

Fig.~\ref{fig:vor_vs_bPara} shows the evolution of four types of vorticities for different collision parameters at 11.5 GeV. At an earlier time, all four types of vorticities are larger at a bigger impact parameter owing to more violent interaction in the more peripheral collision. On the other hand, less colliding matter and shorter evolution also lead to a faster decay at the bigger impact parameter, as the black square line shows. Meanwhile, the ``zero vorticities points'' appear earlier as impact parameters increase.

\subsection{\label{subsec:spactial_vorticity} The spatial distribution of vorticty }
\begin{figure}[htbp]
    %\centering
    \flushleft
    %\flushright
    %\subfigure[NR-vorticity]{
    \subfigure{
        \begin{minipage}[t]{0.5\linewidth}
        %\centering
        %\flushleft
        \includegraphics[scale=0.2]{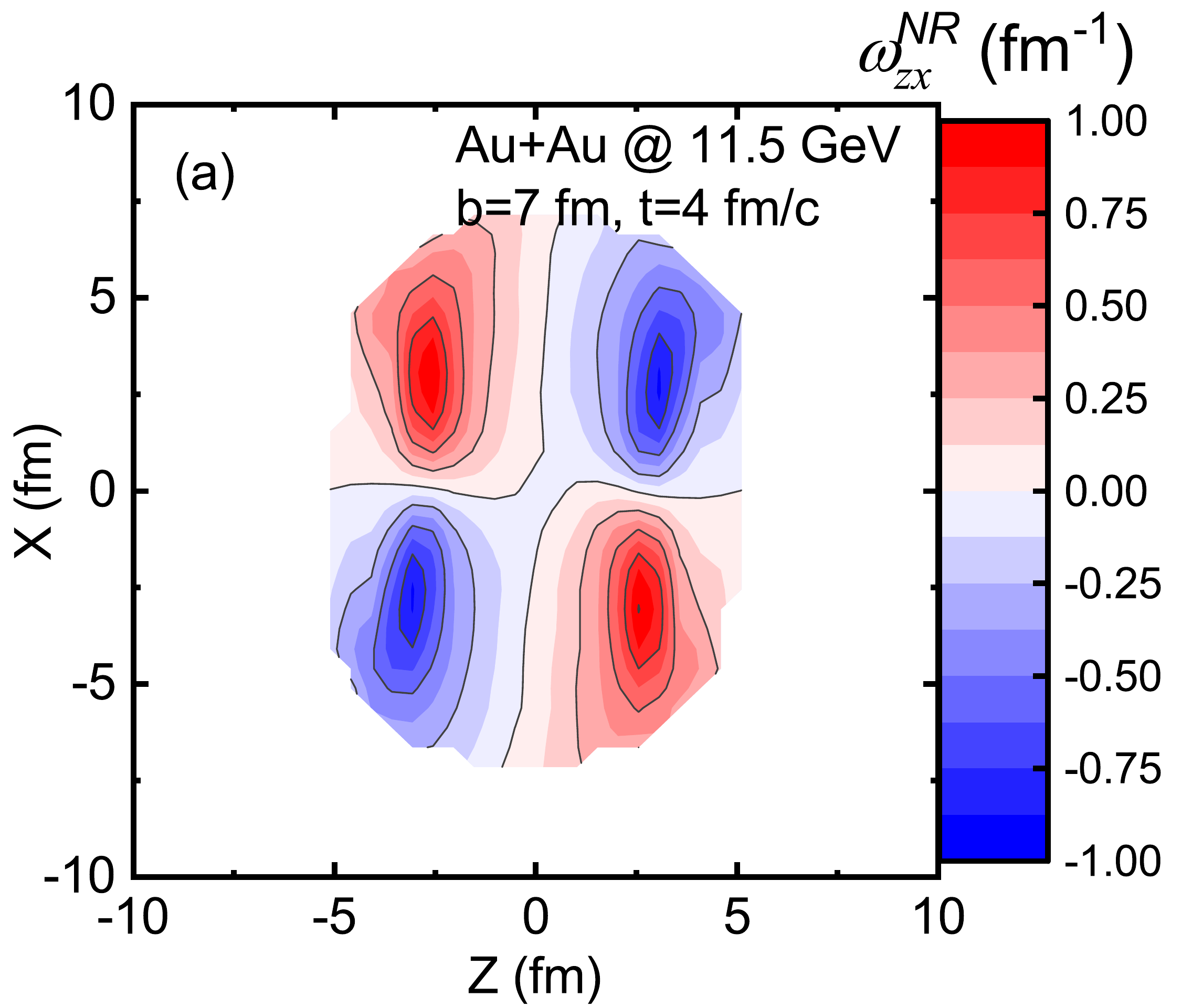}
        %\caption{NR_vor}
        \end{minipage}
        }
    %\subfigure[K-vorticity]{
    \subfigure{
        \begin{minipage}[t]{0\linewidth}
        %\centering
        %\flushright
        \includegraphics[scale=0.2]{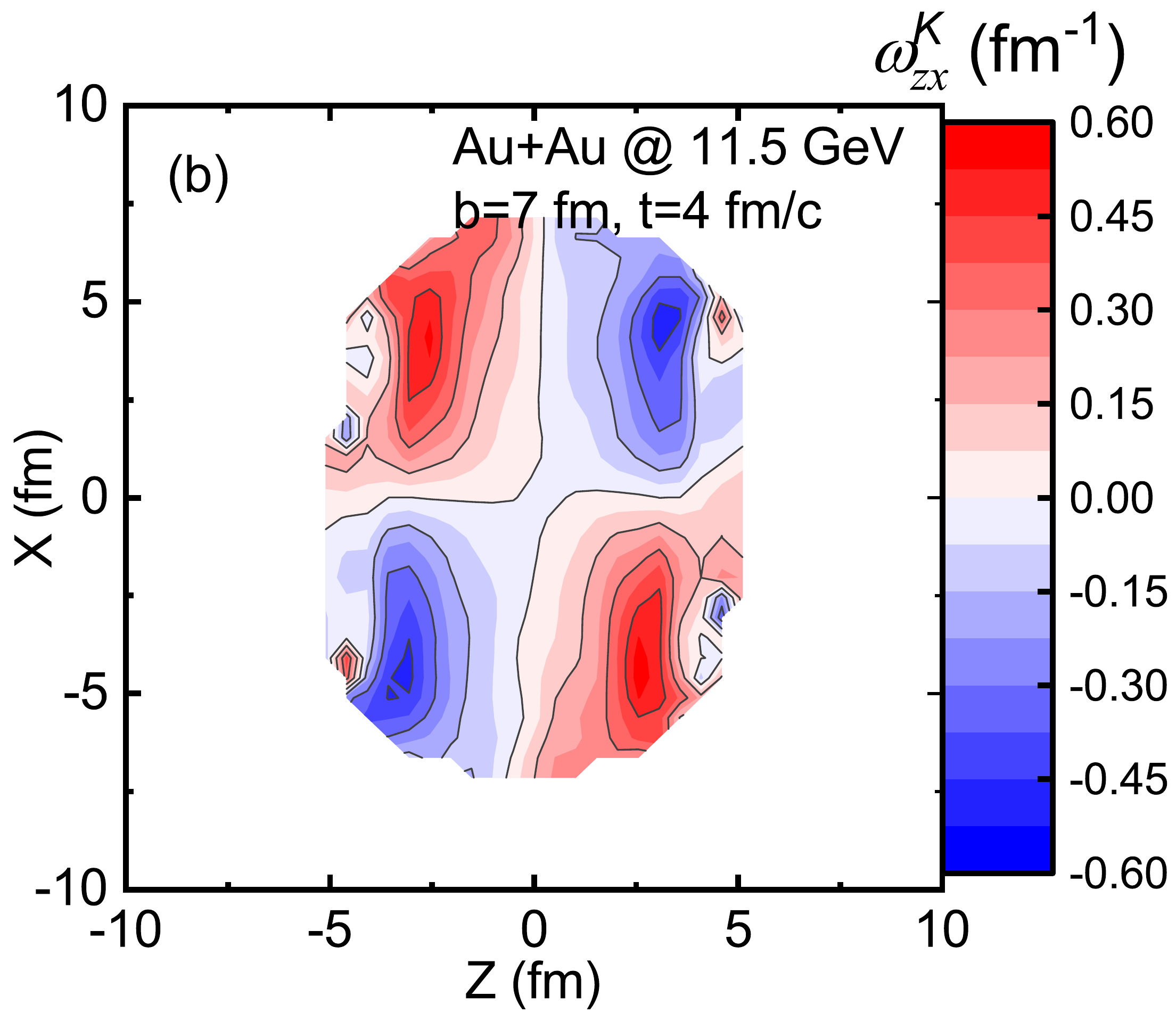}
        %\caption{K_vor}
        \end{minipage}
        } \\
    %\subfigure[T-vorticity]{
    \subfigure{
        \begin{minipage}[t]{0.5\linewidth}
        %\centering
        %\flushleft
        \includegraphics[scale=0.2]{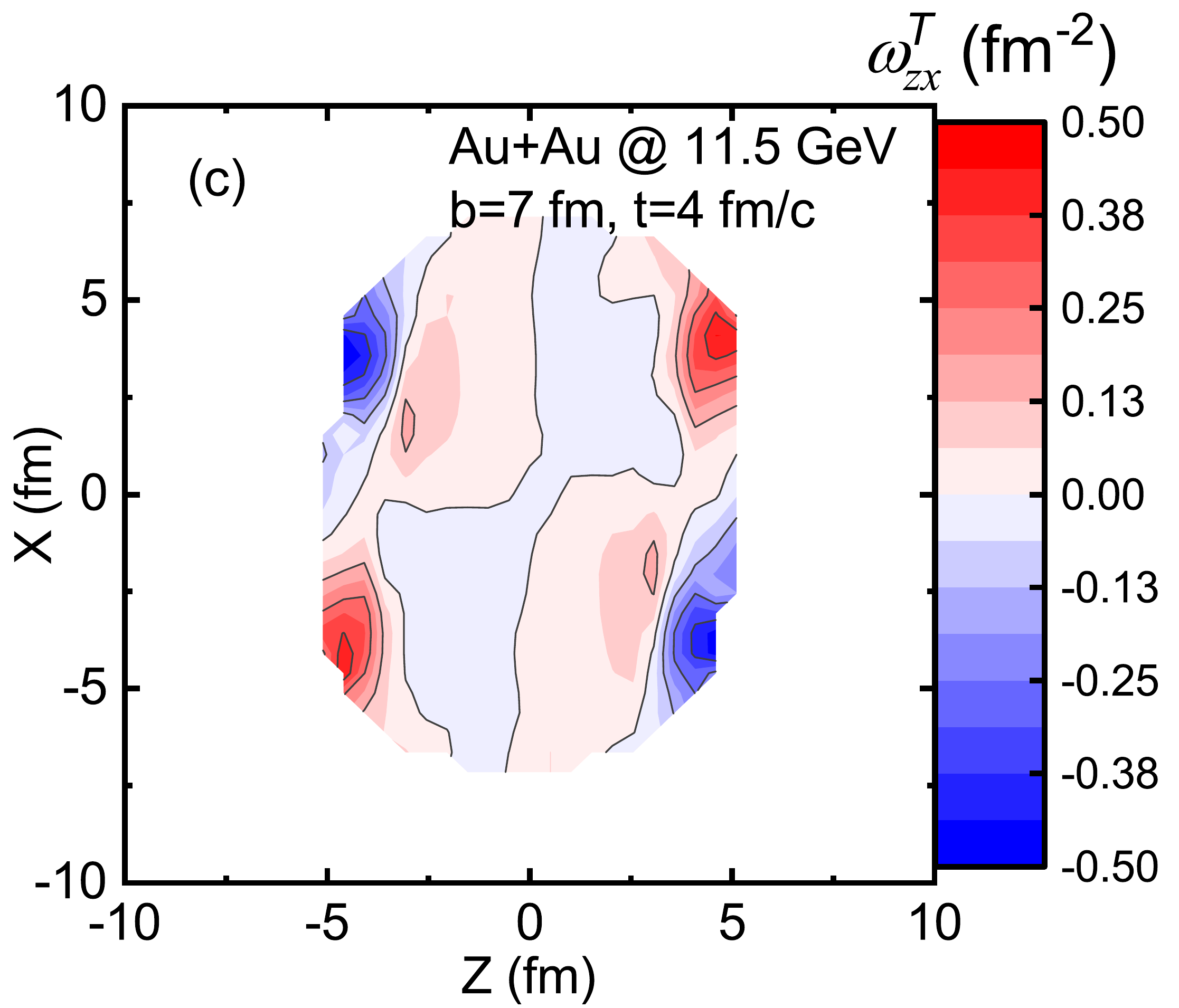}
        %\caption{T_vor}
        \end{minipage}
        }
    %\subfigure[th-vorticity]{
    \subfigure{
        \begin{minipage}[t]{0\linewidth}
        %\centering
        %\flushright
        \includegraphics[scale=0.2]{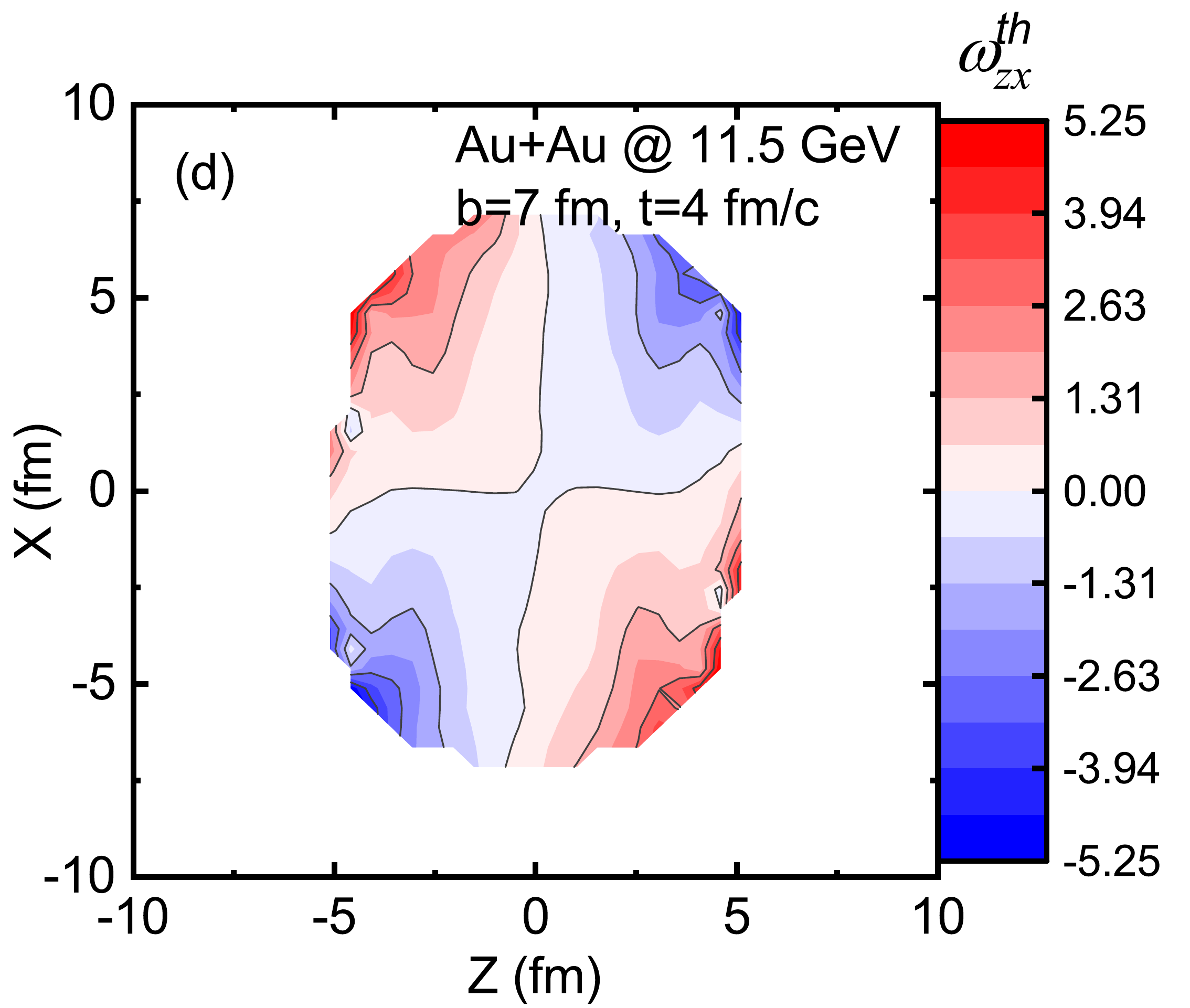}
        %\caption{th_vor}
        \end{minipage}
        }
    %\centering
    \flushleft
    %\flushright
    \caption{\label{fig:vor_spa_dis} The spatial distributions of four types of vorticities in the reaction plane with $b=7$ at $t=$ 4 fm/c for 11.5 GeV in the PACIAE model.}
    \end{figure}

 In Fig.~\ref{fig:vor_spa_dis}, we show the spatial distribution of four types of vorticities in the reaction plane. The pronounced quadrupole structures appear in four types of vorticities. Similar structures have also been found in other studies \cite{Csernai:2013bqa,Teryaev:2015gxa,Jiang:2016woz,Lic:2017sl,Shi:2017wpk,Wei:2018zfb,Vitiuk:2019rfv}. The vorticities are more concentrated in the small areas of the quadrupole structure, that is, the darker parts. Otherwise, there is a small connecting part between the first and fourth quadrants. It means that the overall vorticity will be negative and the direction of the overall particle polarization will point to -Y direction. Compared to the regularly distributed NR-vorticity shown in Fig.~\ref{fig:vor_spa_dis}(a), the K-vorticity has a slight distortion, especially a few areas with opposite signs appear on the border. This comes from the relativistic effect and the discontinuous matter distribution at the edges. For T-vorticity shown in Fig.~\ref{fig:vor_spa_dis}(c), the area of small opposite signs is enlarged because of the direct influence of temperature. However, in the thermal vorticity, such an area is smoothed out and the distribution once again shows a complete quadrupole distribution.

\subsection{\label{subsec:ini_vorticity} Initial vorticty turning point}
\begin{figure}[H]
        %\centering
        \flushleft
        %\flushright
        %\subfigure[NR-vorticity]{
        \subfigure{
            \begin{minipage}[t]{0.5\linewidth}
            %\centering
            %\flushleft
            \includegraphics[scale=0.2]{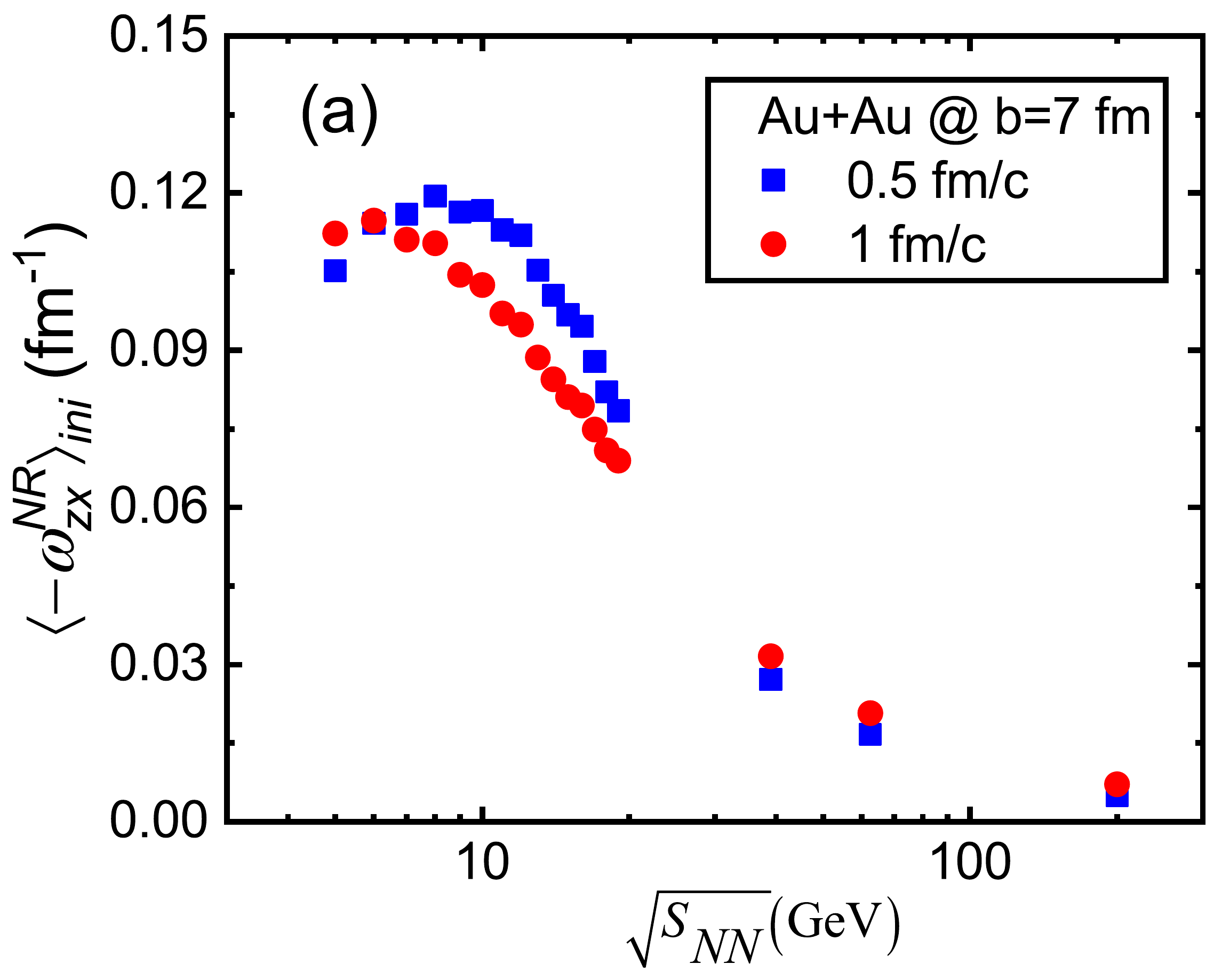}
            %\caption{NR_vor}
            \end{minipage}
            }
        %\subfigure[K-vorticity]{
        \subfigure{
            \begin{minipage}[t]{0\linewidth}
            %\centering
            %\flushright
            \includegraphics[scale=0.2]{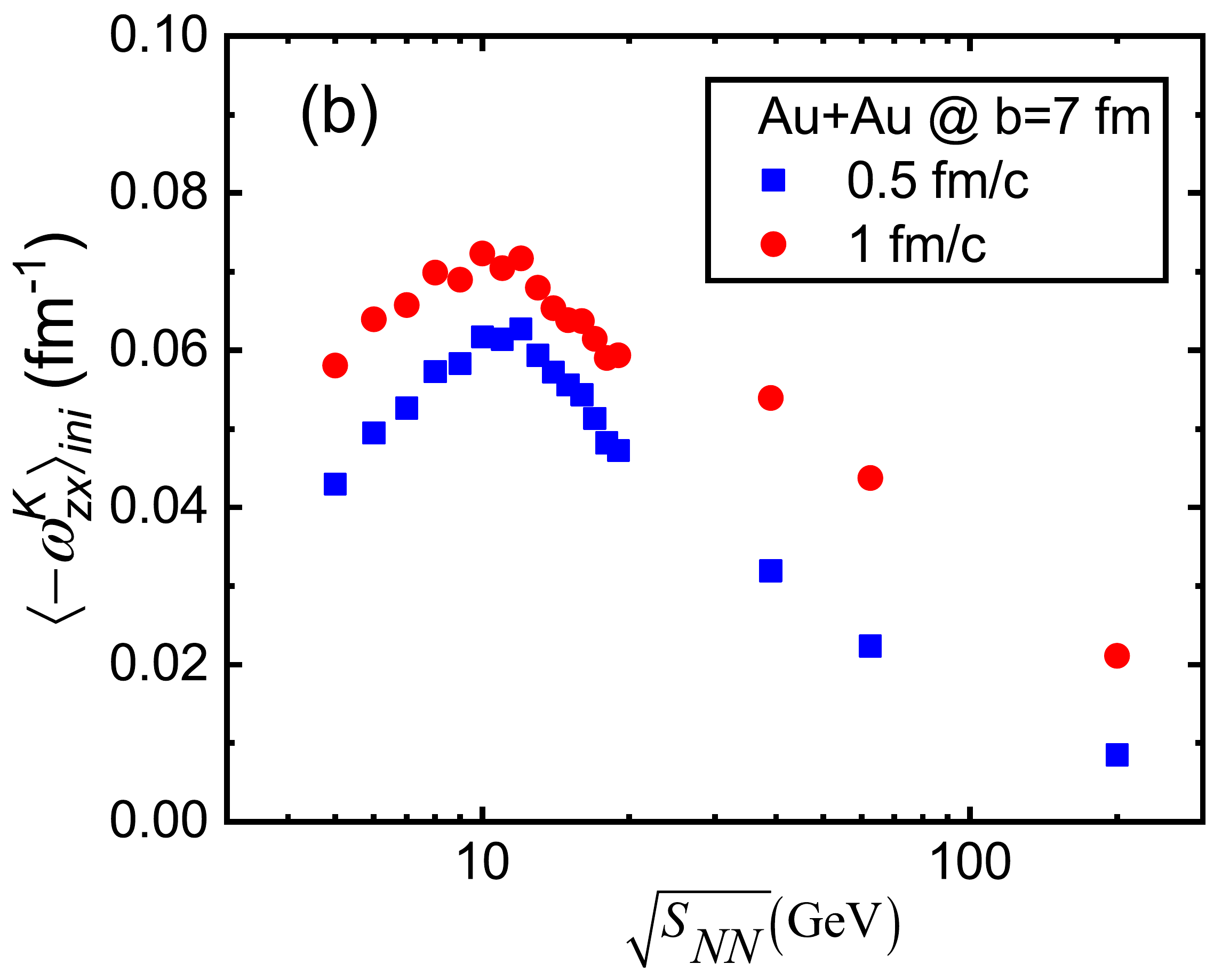}
            %\caption{K_vor}
            \end{minipage}
            } \\
        %\subfigure[T-vorticity]{
        \subfigure{
            \begin{minipage}[t]{0.5\linewidth}
            %\centering
            %\flushleft
            \includegraphics[scale=0.2]{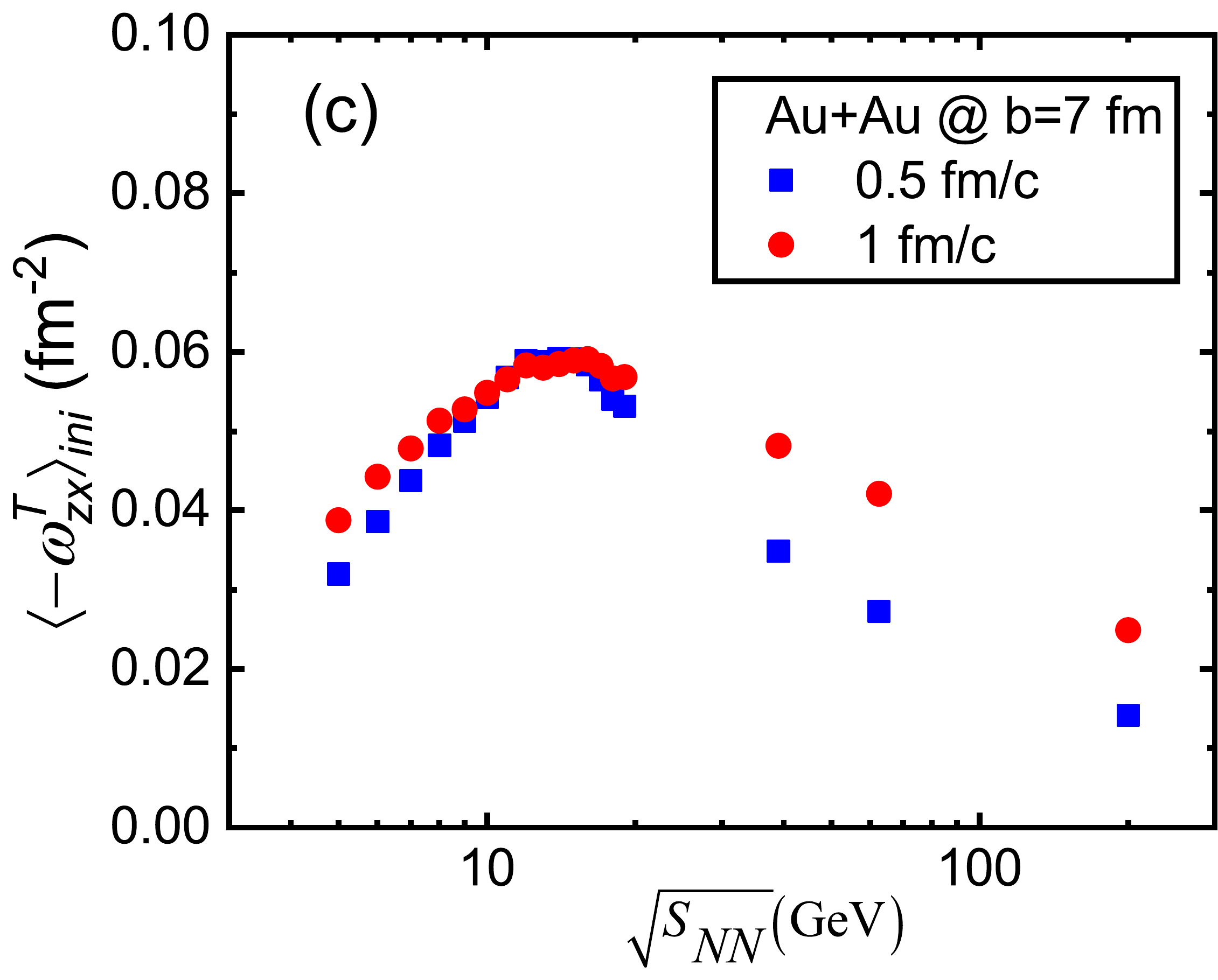}
            %\caption{T_vor}
            \end{minipage}
            }
        %\subfigure[th-vorticity]{
        \subfigure{
            \begin{minipage}[t]{0\linewidth}
            %\centering
            %\flushright
            \includegraphics[scale=0.2]{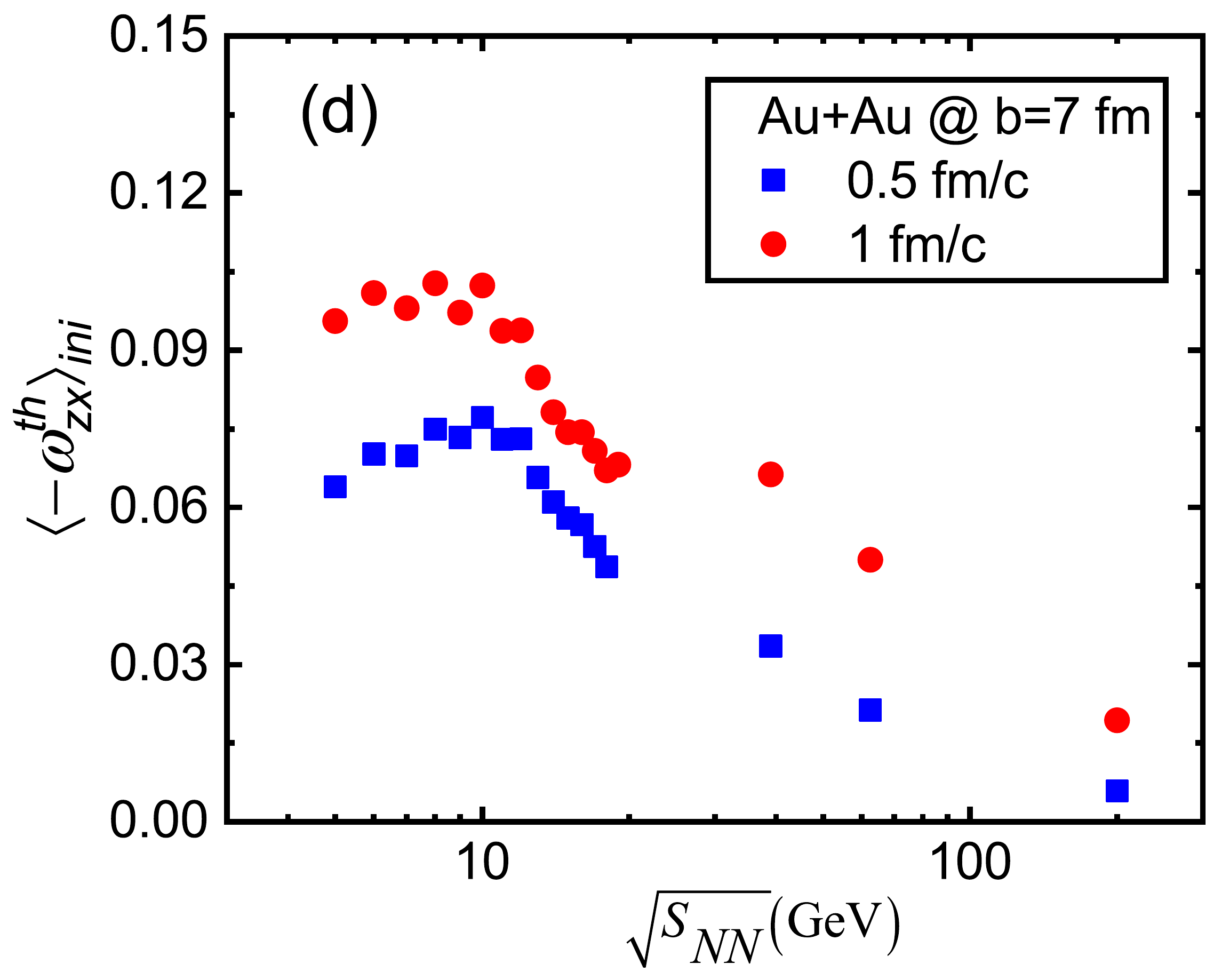}
            %\caption{th_vor}
            \end{minipage}
            }
        %\centering
        \flushleft
        %\flushright
        \caption{\label{fig:iniVor_vs_cmsE} Four types of initial vorticities at $t=$ 0.5 fm/c (blue solid square) and $t=$ 1 fm/c (red solid circle) as function of the collision energy at $b=$ 7 fm .}
    \end{figure}

 In Ref.~\cite{Deng:2020ygd}, it is shown that the so-called initial vorticities first increase then decrease as $\sqrt{S_{NN}}$ grows with a turning point around $\sqrt{S_{NN}} \approx 3-5 \rm GeV$. In Fig.~\ref{fig:iniVor_vs_cmsE}, we show the values of the initial vorticities among four types of vorticities at 0.5 fm/c (blue solid square) and 1 fm/c (red solid circle) which could be compared with work in Ref.~\cite{Deng:2020ygd}. Our results show such a turning point at around 10 GeV in NR-vorticity, K-vorticity and thermal vorticity. In Fig.~\ref{fig:iniVor_vs_cmsE}(c), the turning point of T-vorticity moves back to around 15 GeV under the influence of temperature. Although our turning energy is different from the previous study in UrQMD and IQMD \cite{Deng:2020ygd}, 
our results do show that the initial vorticities have a non-monotonic dependence on the energy. This implies the non-monotonicity of polarization to collision energies, which needs to be verified by future experiments. For lower collision energies, the PACIAE model is no longer applicable, which requires the use of other suitable models.

\subsection{\label{subsec:polarization} $\Lambda$ polarization in PACIAE }

As we mentioned in section \ref{sec:model}, PACIAE does not implement a phase transition. In order to deal with the situation, here we introduce a simple freeze-out criterion that the parton system hadronizes when the cell-average energy density is reduced to about 0.1 ${\rm GeV/fm^{3}}$. On the other hand, the formation of hadrons is assumed to be instantaneous. The hadrons are produced and emitted from the parton matter. Hence, the temperature and related thermal voticitity will be calculated based on the partons at the moment of hadrons formation following Eq.~(\ref{Eq:temp_parton}) and (\ref{Eq:th_vor_tens}), respectively. The produced $\Lambda$ hyperons will be used to calculate spin-polarization following Eq.~(\ref{Eq:spin_pol_orig}-\ref{Eq:spin_pol_glob}).
    \begin{figure}[h]
        \centering
        \includegraphics[scale=0.45,width=0.45\textwidth]{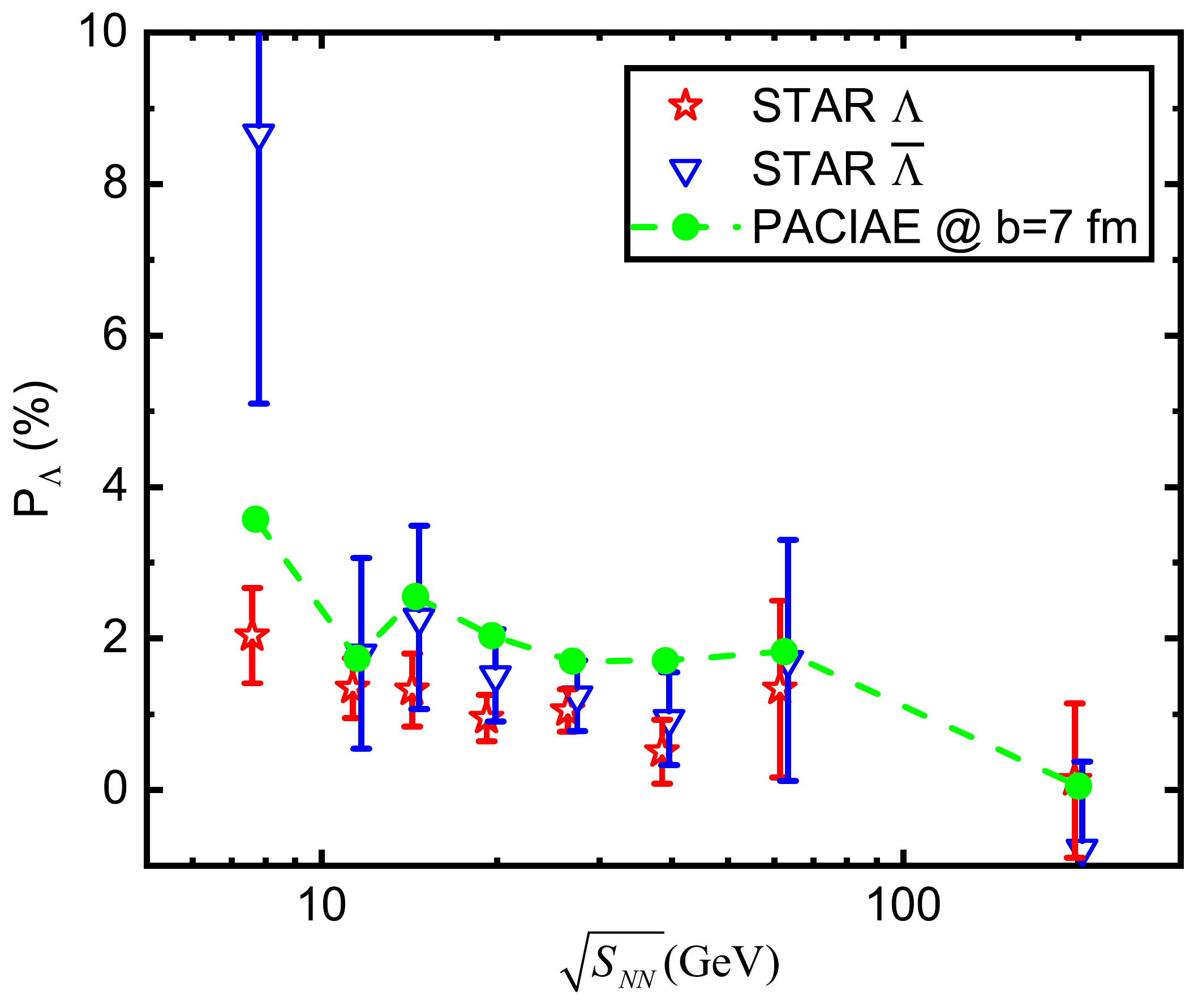}
        \caption{The gobal $\Lambda$ polarization of Au+Au collision as a function of collision energy with b=7 fm in PACIAE compared with STAR experimental data \cite{STAR:2017ckg}(red open star and blue open inverted triangle). The results of this work are shown by green solid circles.}
        \label{fig:gloPolVsCmsE}
    \end{figure}
    
    In Fig.~\ref{fig:gloPolVsCmsE}, we show our calculation results of global $\Lambda$ polarization for Au+Au collisions at $b=7$ fm corresponding to the $ 20\%-50\% $ centrality bin of STAR experiments. The polarization results of this work are only based on the default input parameters of the PACIAE model and do not consider the feed-down effect from higher-lying resonances. However, our results well reproduce the STAR experimental data at energies 7.7, 11.5, 14.5, 19.6, 27, 39, 62.4 and 200 GeV \cite{STAR:2017ckg}. Our results also show that the $\Lambda$ polarization decreases with the increase of collision energy, except for the slight deviation point of 11.5 GeV. 

\section{\label{sec:summary} SUMMARY }
    We calculated the non-relativistic, kinematic, temperature and thermal vorticities at energies $ \sqrt{S_{NN}} $ = 5--200 GeV employing the PACIAE model. The energy, time and centrality dependence of four types of vorticities have been studied. A faster decay at lower energy and smaller impact parameters was found. The behaviors are consistent with our understanding of heavy-ion collisions. The quadrupole structures of spatial distribution in the reaction plane have been found among four types of vorticities.
    
    It was reconfirmed that the initial vorticities have a non-monotonous dependence on increasing collision energies, and the turning point was 10-15 GeV. Such behavior may bring about non-monotonic dependence of particle polarization on energies. It requires future verification and researches in lower energy experiments.
    
    The $\Lambda$ polarization for Au+Au collisions is calculated at STAR BES energies $ \sqrt{S_{NN}} $ = 7.7--200 GeV. The results agree with experimental measurements. This is an extension of the PACIAE model to polarization research. More polarization investigations in PACIAE, such as local polarization, the polarization of different particles and different collision systems, will be performed in our future works.
\begin{acknowledgments}
    The authors thank Yilong Xie and Larissa V. Bravina for helpful discussions. This work was supported by National Natural Science Foundation of China (Grant No. 11905163, 11775094)  %TODO
\end{acknowledgments}

\normalem   %Lei20211015 without underline

\bibliography{Vor_Pol_in_PACIAE}

\begin{thebibliography}{49}%
\makeatletter
\providecommand \@ifxundefined [1]{%
 \@ifx{#1\undefined}
}%
\providecommand \@ifnum [1]{%
 \ifnum #1\expandafter \@firstoftwo
 \else \expandafter \@secondoftwo
 \fi
}%
\providecommand \@ifx [1]{%
 \ifx #1\expandafter \@firstoftwo
 \else \expandafter \@secondoftwo
 \fi
}%
\providecommand \natexlab [1]{#1}%
\providecommand \enquote  [1]{``#1''}%
\providecommand \bibnamefont  [1]{#1}%
\providecommand \bibfnamefont [1]{#1}%
\providecommand \citenamefont [1]{#1}%
\providecommand \href@noop [0]{\@secondoftwo}%
\providecommand \href [0]{\begingroup \@sanitize@url \@href}%
\providecommand \@href[1]{\@@startlink{#1}\@@href}%
\providecommand \@@href[1]{\endgroup#1\@@endlink}%
\providecommand \@sanitize@url [0]{\catcode `\\12\catcode `\$12\catcode
  `\&12\catcode `\#12\catcode `\^12\catcode `\_12\catcode `\%12\relax}%
\providecommand \@@startlink[1]{}%
\providecommand \@@endlink[0]{}%
\providecommand \url  [0]{\begingroup\@sanitize@url \@url }%
\providecommand \@url [1]{\endgroup\@href {#1}{\urlprefix }}%
\providecommand \urlprefix  [0]{URL }%
\providecommand \Eprint [0]{\href }%
\providecommand \doibase [0]{https://doi.org/}%
\providecommand \selectlanguage [0]{\@gobble}%
\providecommand \bibinfo  [0]{\@secondoftwo}%
\providecommand \bibfield  [0]{\@secondoftwo}%
\providecommand \translation [1]{[#1]}%
\providecommand \BibitemOpen [0]{}%
\providecommand \bibitemStop [0]{}%
\providecommand \bibitemNoStop [0]{.\EOS\space}%
\providecommand \EOS [0]{\spacefactor3000\relax}%
\providecommand \BibitemShut  [1]{\csname bibitem#1\endcsname}%
\let\auto@bib@innerbib\@empty
%</preamble>
\bibitem [{\citenamefont {Liang}\ and\ \citenamefont
  {Wang}(2005)}]{Liang:2004ph}%
  \BibitemOpen
  \bibfield  {author} {\bibinfo {author} {\bibfnamefont {Z.-T.}\ \bibnamefont
  {Liang}}\ and\ \bibinfo {author} {\bibfnamefont {X.-N.}\ \bibnamefont
  {Wang}},\ }\href {https://doi.org/10.1103/PhysRevLett.94.102301} {\bibfield
  {journal} {\bibinfo  {journal} {Phys. Rev. Lett.}\ }\textbf {\bibinfo
  {volume} {94}},\ \bibinfo {pages} {102301} (\bibinfo {year} {2005})},\
  \bibinfo {note} {[Erratum: Phys.Rev.Lett. 96, 039901 (2006)]},\ \Eprint
  {https://arxiv.org/abs/nucl-th/0410079} {arXiv:nucl-th/0410079} \BibitemShut
  {NoStop}%
\bibitem [{\citenamefont {Voloshin}()}]{Voloshin:2004ha}%
  \BibitemOpen
  \bibfield  {author} {\bibinfo {author} {\bibfnamefont {S.~A.}\ \bibnamefont
  {Voloshin}},\ }\href@noop {} {}\Eprint
  {https://arxiv.org/abs/nucl-th/0410089} {arXiv:nucl-th/0410089} \BibitemShut
  {NoStop}%
\bibitem [{\citenamefont {Becattini}\ and\ \citenamefont
  {Piccinini}(2008)}]{Becattini:2007nd}%
  \BibitemOpen
  \bibfield  {author} {\bibinfo {author} {\bibfnamefont {F.}~\bibnamefont
  {Becattini}}\ and\ \bibinfo {author} {\bibfnamefont {F.}~\bibnamefont
  {Piccinini}},\ }\href {https://doi.org/10.1016/j.aop.2008.01.001} {\bibfield
  {journal} {\bibinfo  {journal} {Annals Phys.}\ }\textbf {\bibinfo {volume}
  {323}},\ \bibinfo {pages} {2452} (\bibinfo {year} {2008})},\ \Eprint
  {https://arxiv.org/abs/0710.5694} {arXiv:0710.5694 [nucl-th]} \BibitemShut
  {NoStop}%
\bibitem [{\citenamefont {Becattini}\ \emph {et~al.}(2008)\citenamefont
  {Becattini}, \citenamefont {Piccinini},\ and\ \citenamefont
  {Rizzo}}]{Becattini:2007sr}%
  \BibitemOpen
  \bibfield  {author} {\bibinfo {author} {\bibfnamefont {F.}~\bibnamefont
  {Becattini}}, \bibinfo {author} {\bibfnamefont {F.}~\bibnamefont
  {Piccinini}},\ and\ \bibinfo {author} {\bibfnamefont {J.}~\bibnamefont
  {Rizzo}},\ }\href {https://doi.org/10.1103/PhysRevC.77.024906} {\bibfield
  {journal} {\bibinfo  {journal} {Phys. Rev. C}\ }\textbf {\bibinfo {volume}
  {77}},\ \bibinfo {pages} {024906} (\bibinfo {year} {2008})},\ \Eprint
  {https://arxiv.org/abs/0711.1253} {arXiv:0711.1253 [nucl-th]} \BibitemShut
  {NoStop}%
\bibitem [{\citenamefont {Becattini}\ \emph
  {et~al.}(2013{\natexlab{a}})\citenamefont {Becattini}, \citenamefont
  {Chandra}, \citenamefont {Del~Zanna},\ and\ \citenamefont
  {Grossi}}]{Becattini:2013fla}%
  \BibitemOpen
  \bibfield  {author} {\bibinfo {author} {\bibfnamefont {F.}~\bibnamefont
  {Becattini}}, \bibinfo {author} {\bibfnamefont {V.}~\bibnamefont {Chandra}},
  \bibinfo {author} {\bibfnamefont {L.}~\bibnamefont {Del~Zanna}},\ and\
  \bibinfo {author} {\bibfnamefont {E.}~\bibnamefont {Grossi}},\ }\href
  {https://doi.org/10.1016/j.aop.2013.07.004} {\bibfield  {journal} {\bibinfo
  {journal} {Annals Phys.}\ }\textbf {\bibinfo {volume} {338}},\ \bibinfo
  {pages} {32} (\bibinfo {year} {2013}{\natexlab{a}})},\ \Eprint
  {https://arxiv.org/abs/1303.3431} {arXiv:1303.3431 [nucl-th]} \BibitemShut
  {NoStop}%
\bibitem [{\citenamefont {Fang}\ \emph {et~al.}(2016)\citenamefont {Fang},
  \citenamefont {Pang}, \citenamefont {Wang},\ and\ \citenamefont
  {Wang}}]{Fang:2016vpj}%
  \BibitemOpen
  \bibfield  {author} {\bibinfo {author} {\bibfnamefont {R.-H.}\ \bibnamefont
  {Fang}}, \bibinfo {author} {\bibfnamefont {L.-G.}\ \bibnamefont {Pang}},
  \bibinfo {author} {\bibfnamefont {Q.}~\bibnamefont {Wang}},\ and\ \bibinfo
  {author} {\bibfnamefont {X.-N.}\ \bibnamefont {Wang}},\ }\href
  {https://doi.org/10.1103/PhysRevC.94.024904} {\bibfield  {journal} {\bibinfo
  {journal} {Phys. Rev. C}\ }\textbf {\bibinfo {volume} {94}},\ \bibinfo
  {pages} {024904} (\bibinfo {year} {2016})},\ \Eprint
  {https://arxiv.org/abs/1604.04036} {arXiv:1604.04036 [nucl-th]} \BibitemShut
  {NoStop}%
\bibitem [{\citenamefont {Csernai}\ \emph {et~al.}(2013)\citenamefont
  {Csernai}, \citenamefont {Magas},\ and\ \citenamefont
  {Wang}}]{Csernai:2013bqa}%
  \BibitemOpen
  \bibfield  {author} {\bibinfo {author} {\bibfnamefont {L.~P.}\ \bibnamefont
  {Csernai}}, \bibinfo {author} {\bibfnamefont {V.~K.}\ \bibnamefont {Magas}},\
  and\ \bibinfo {author} {\bibfnamefont {D.~J.}\ \bibnamefont {Wang}},\ }\href
  {https://doi.org/10.1103/PhysRevC.87.034906} {\bibfield  {journal} {\bibinfo
  {journal} {Phys. Rev. C}\ }\textbf {\bibinfo {volume} {87}},\ \bibinfo
  {pages} {034906} (\bibinfo {year} {2013})},\ \Eprint
  {https://arxiv.org/abs/1302.5310} {arXiv:1302.5310 [nucl-th]} \BibitemShut
  {NoStop}%
\bibitem [{\citenamefont {Becattini}\ \emph
  {et~al.}(2013{\natexlab{b}})\citenamefont {Becattini}, \citenamefont
  {Csernai},\ and\ \citenamefont {Wang}}]{Becattini:2013vja}%
  \BibitemOpen
  \bibfield  {author} {\bibinfo {author} {\bibfnamefont {F.}~\bibnamefont
  {Becattini}}, \bibinfo {author} {\bibfnamefont {L.}~\bibnamefont {Csernai}},\
  and\ \bibinfo {author} {\bibfnamefont {D.~J.}\ \bibnamefont {Wang}},\ }\href
  {https://doi.org/10.1103/PhysRevC.88.034905} {\bibfield  {journal} {\bibinfo
  {journal} {Phys. Rev. C}\ }\textbf {\bibinfo {volume} {88}},\ \bibinfo
  {pages} {034905} (\bibinfo {year} {2013}{\natexlab{b}})},\ \bibinfo {note}
  {[Erratum: Phys.Rev.C 93, 069901 (2016)]},\ \Eprint
  {https://arxiv.org/abs/1304.4427} {arXiv:1304.4427 [nucl-th]} \BibitemShut
  {NoStop}%
\bibitem [{\citenamefont {Xie}\ \emph {et~al.}(2016)\citenamefont {Xie},
  \citenamefont {Bleicher}, \citenamefont {St\"ocker}, \citenamefont {Wang},\
  and\ \citenamefont {Csernai}}]{Xie:2016fjj}%
  \BibitemOpen
  \bibfield  {author} {\bibinfo {author} {\bibfnamefont {Y.~L.}\ \bibnamefont
  {Xie}}, \bibinfo {author} {\bibfnamefont {M.}~\bibnamefont {Bleicher}},
  \bibinfo {author} {\bibfnamefont {H.}~\bibnamefont {St\"ocker}}, \bibinfo
  {author} {\bibfnamefont {D.~J.}\ \bibnamefont {Wang}},\ and\ \bibinfo
  {author} {\bibfnamefont {L.~P.}\ \bibnamefont {Csernai}},\ }\href
  {https://doi.org/10.1103/PhysRevC.94.054907} {\bibfield  {journal} {\bibinfo
  {journal} {Phys. Rev. C}\ }\textbf {\bibinfo {volume} {94}},\ \bibinfo
  {pages} {054907} (\bibinfo {year} {2016})},\ \Eprint
  {https://arxiv.org/abs/1610.08678} {arXiv:1610.08678 [nucl-th]} \BibitemShut
  {NoStop}%
\bibitem [{\citenamefont {Xie}\ \emph {et~al.}(2017)\citenamefont {Xie},
  \citenamefont {Wang},\ and\ \citenamefont {Csernai}}]{Xie:2017upb}%
  \BibitemOpen
  \bibfield  {author} {\bibinfo {author} {\bibfnamefont {Y.}~\bibnamefont
  {Xie}}, \bibinfo {author} {\bibfnamefont {D.}~\bibnamefont {Wang}},\ and\
  \bibinfo {author} {\bibfnamefont {L.~P.}\ \bibnamefont {Csernai}},\ }\href
  {https://doi.org/10.1103/PhysRevC.95.031901} {\bibfield  {journal} {\bibinfo
  {journal} {Phys. Rev. C}\ }\textbf {\bibinfo {volume} {95}},\ \bibinfo
  {pages} {031901} (\bibinfo {year} {2017})},\ \Eprint
  {https://arxiv.org/abs/1703.03770} {arXiv:1703.03770 [nucl-th]} \BibitemShut
  {NoStop}%
\bibitem [{\citenamefont {Xie}\ \emph {et~al.}(2021)\citenamefont {Xie},
  \citenamefont {Chen},\ and\ \citenamefont {Csernai}}]{Xie:2021fjn}%
  \BibitemOpen
  \bibfield  {author} {\bibinfo {author} {\bibfnamefont {Y.}~\bibnamefont
  {Xie}}, \bibinfo {author} {\bibfnamefont {G.}~\bibnamefont {Chen}},\ and\
  \bibinfo {author} {\bibfnamefont {L.~P.}\ \bibnamefont {Csernai}},\ }\href
  {https://doi.org/10.1140/epjc/s10052-021-08828-z} {\bibfield  {journal}
  {\bibinfo  {journal} {Eur. Phys. J. C}\ }\textbf {\bibinfo {volume} {81}},\
  \bibinfo {pages} {12} (\bibinfo {year} {2021})},\ \Eprint
  {https://arxiv.org/abs/1912.00209} {arXiv:1912.00209 [hep-ph]} \BibitemShut
  {NoStop}%
\bibitem [{\citenamefont {Karpenko}\ and\ \citenamefont
  {Becattini}(2017{\natexlab{a}})}]{Karpenko:2016jyx}%
  \BibitemOpen
  \bibfield  {author} {\bibinfo {author} {\bibfnamefont {I.}~\bibnamefont
  {Karpenko}}\ and\ \bibinfo {author} {\bibfnamefont {F.}~\bibnamefont
  {Becattini}},\ }\href {https://doi.org/10.1140/epjc/s10052-017-4765-1}
  {\bibfield  {journal} {\bibinfo  {journal} {Eur. Phys. J. C}\ }\textbf
  {\bibinfo {volume} {77}},\ \bibinfo {pages} {213} (\bibinfo {year}
  {2017}{\natexlab{a}})},\ \Eprint {https://arxiv.org/abs/1610.04717}
  {arXiv:1610.04717 [nucl-th]} \BibitemShut {NoStop}%
\bibitem [{\citenamefont {Becattini}\ \emph {et~al.}()\citenamefont
  {Becattini}, \citenamefont {Buzzegoli}, \citenamefont {Palermo},
  \citenamefont {Inghirami},\ and\ \citenamefont
  {Karpenko}}]{Becattini:2021iol}%
  \BibitemOpen
  \bibfield  {author} {\bibinfo {author} {\bibfnamefont {F.}~\bibnamefont
  {Becattini}}, \bibinfo {author} {\bibfnamefont {M.}~\bibnamefont
  {Buzzegoli}}, \bibinfo {author} {\bibfnamefont {A.}~\bibnamefont {Palermo}},
  \bibinfo {author} {\bibfnamefont {G.}~\bibnamefont {Inghirami}},\ and\
  \bibinfo {author} {\bibfnamefont {I.}~\bibnamefont {Karpenko}},\ }\href@noop
  {} {}\Eprint {https://arxiv.org/abs/2103.14621} {arXiv:2103.14621 [nucl-th]}
  \BibitemShut {NoStop}%
\bibitem [{\citenamefont {Ivanov}\ \emph {et~al.}(2019)\citenamefont {Ivanov},
  \citenamefont {Toneev},\ and\ \citenamefont
  {Soldatov}}]{PhysRevC.100.014908}%
  \BibitemOpen
  \bibfield  {author} {\bibinfo {author} {\bibfnamefont {Y.~B.}\ \bibnamefont
  {Ivanov}}, \bibinfo {author} {\bibfnamefont {V.~D.}\ \bibnamefont {Toneev}},\
  and\ \bibinfo {author} {\bibfnamefont {A.~A.}\ \bibnamefont {Soldatov}},\
  }\href {https://doi.org/10.1103/PhysRevC.100.014908} {\bibfield  {journal}
  {\bibinfo  {journal} {Phys. Rev. C}\ }\textbf {\bibinfo {volume} {100}},\
  \bibinfo {pages} {014908} (\bibinfo {year} {2019})}\BibitemShut {NoStop}%
\bibitem [{\citenamefont {Ivanov}\ and\ \citenamefont
  {Soldatov}(2020)}]{Ivanov:2020wak}%
  \BibitemOpen
  \bibfield  {author} {\bibinfo {author} {\bibfnamefont {Y.~B.}\ \bibnamefont
  {Ivanov}}\ and\ \bibinfo {author} {\bibfnamefont {A.~A.}\ \bibnamefont
  {Soldatov}},\ }\href {https://doi.org/10.1103/PhysRevC.102.024916} {\bibfield
   {journal} {\bibinfo  {journal} {Phys. Rev. C}\ }\textbf {\bibinfo {volume}
  {102}},\ \bibinfo {pages} {024916} (\bibinfo {year} {2020})},\ \Eprint
  {https://arxiv.org/abs/2004.05166} {arXiv:2004.05166 [nucl-th]} \BibitemShut
  {NoStop}%
\bibitem [{\citenamefont {Jiang}\ \emph {et~al.}(2016)\citenamefont {Jiang},
  \citenamefont {Lin},\ and\ \citenamefont {Liao}}]{Jiang:2016woz}%
  \BibitemOpen
  \bibfield  {author} {\bibinfo {author} {\bibfnamefont {Y.}~\bibnamefont
  {Jiang}}, \bibinfo {author} {\bibfnamefont {Z.-W.}\ \bibnamefont {Lin}},\
  and\ \bibinfo {author} {\bibfnamefont {J.}~\bibnamefont {Liao}},\ }\href
  {https://doi.org/10.1103/PhysRevC.94.044910} {\bibfield  {journal} {\bibinfo
  {journal} {Phys. Rev. C}\ }\textbf {\bibinfo {volume} {94}},\ \bibinfo
  {pages} {044910} (\bibinfo {year} {2016})},\ \bibinfo {note} {[Erratum:
  Phys.Rev.C 95, 049904 (2017)]},\ \Eprint {https://arxiv.org/abs/1602.06580}
  {arXiv:1602.06580 [hep-ph]} \BibitemShut {NoStop}%
\bibitem [{\citenamefont {Li}\ \emph {et~al.}(2017)\citenamefont {Li},
  \citenamefont {Pang}, \citenamefont {Wang},\ and\ \citenamefont
  {Xia}}]{Lic:2017sl}%
  \BibitemOpen
  \bibfield  {author} {\bibinfo {author} {\bibfnamefont {H.}~\bibnamefont
  {Li}}, \bibinfo {author} {\bibfnamefont {L.-G.}\ \bibnamefont {Pang}},
  \bibinfo {author} {\bibfnamefont {Q.}~\bibnamefont {Wang}},\ and\ \bibinfo
  {author} {\bibfnamefont {X.-L.}\ \bibnamefont {Xia}},\ }\href
  {https://doi.org/10.1103/PhysRevC.96.054908} {\bibfield  {journal} {\bibinfo
  {journal} {Phys. Rev. C}\ }\textbf {\bibinfo {volume} {96}},\ \bibinfo
  {pages} {054908} (\bibinfo {year} {2017})},\ \Eprint
  {https://arxiv.org/abs/1704.01507} {arXiv:1704.01507 [nucl-th]} \BibitemShut
  {NoStop}%
\bibitem [{\citenamefont {Xia}\ \emph {et~al.}(2018)\citenamefont {Xia},
  \citenamefont {Li}, \citenamefont {Tang},\ and\ \citenamefont
  {Wang}}]{Xia:2018tes}%
  \BibitemOpen
  \bibfield  {author} {\bibinfo {author} {\bibfnamefont {X.-L.}\ \bibnamefont
  {Xia}}, \bibinfo {author} {\bibfnamefont {H.}~\bibnamefont {Li}}, \bibinfo
  {author} {\bibfnamefont {Z.-B.}\ \bibnamefont {Tang}},\ and\ \bibinfo
  {author} {\bibfnamefont {Q.}~\bibnamefont {Wang}},\ }\href
  {https://doi.org/10.1103/PhysRevC.98.024905} {\bibfield  {journal} {\bibinfo
  {journal} {Phys. Rev. C}\ }\textbf {\bibinfo {volume} {98}},\ \bibinfo
  {pages} {024905} (\bibinfo {year} {2018})},\ \Eprint
  {https://arxiv.org/abs/1803.00867} {arXiv:1803.00867 [nucl-th]} \BibitemShut
  {NoStop}%
\bibitem [{\citenamefont {Shi}\ \emph {et~al.}(2019)\citenamefont {Shi},
  \citenamefont {Li},\ and\ \citenamefont {Liao}}]{Shi:2017wpk}%
  \BibitemOpen
  \bibfield  {author} {\bibinfo {author} {\bibfnamefont {S.}~\bibnamefont
  {Shi}}, \bibinfo {author} {\bibfnamefont {K.}~\bibnamefont {Li}},\ and\
  \bibinfo {author} {\bibfnamefont {J.}~\bibnamefont {Liao}},\ }\href
  {https://doi.org/10.1016/j.physletb.2018.09.066} {\bibfield  {journal}
  {\bibinfo  {journal} {Phys. Lett. B}\ }\textbf {\bibinfo {volume} {788}},\
  \bibinfo {pages} {409} (\bibinfo {year} {2019})},\ \Eprint
  {https://arxiv.org/abs/1712.00878} {arXiv:1712.00878 [nucl-th]} \BibitemShut
  {NoStop}%
\bibitem [{\citenamefont {Wei}\ \emph {et~al.}(2019)\citenamefont {Wei},
  \citenamefont {Deng},\ and\ \citenamefont {Huang}}]{Wei:2018zfb}%
  \BibitemOpen
  \bibfield  {author} {\bibinfo {author} {\bibfnamefont {D.-X.}\ \bibnamefont
  {Wei}}, \bibinfo {author} {\bibfnamefont {W.-T.}\ \bibnamefont {Deng}},\ and\
  \bibinfo {author} {\bibfnamefont {X.-G.}\ \bibnamefont {Huang}},\ }\href
  {https://doi.org/10.1103/PhysRevC.99.014905} {\bibfield  {journal} {\bibinfo
  {journal} {Phys. Rev. C}\ }\textbf {\bibinfo {volume} {99}},\ \bibinfo
  {pages} {014905} (\bibinfo {year} {2019})},\ \Eprint
  {https://arxiv.org/abs/1810.00151} {arXiv:1810.00151 [nucl-th]} \BibitemShut
  {NoStop}%
\bibitem [{\citenamefont {Karpenko}\ and\ \citenamefont
  {Becattini}(2017{\natexlab{b}})}]{Karpenko:2017lyj}%
  \BibitemOpen
  \bibfield  {author} {\bibinfo {author} {\bibfnamefont {I.}~\bibnamefont
  {Karpenko}}\ and\ \bibinfo {author} {\bibfnamefont {F.}~\bibnamefont
  {Becattini}},\ }\href {https://doi.org/10.1016/j.nuclphysa.2017.05.057}
  {\bibfield  {journal} {\bibinfo  {journal} {Nucl. Phys. A}\ }\textbf
  {\bibinfo {volume} {967}},\ \bibinfo {pages} {764} (\bibinfo {year}
  {2017}{\natexlab{b}})},\ \Eprint {https://arxiv.org/abs/1704.02142}
  {arXiv:1704.02142 [nucl-th]} \BibitemShut {NoStop}%
\bibitem [{\citenamefont {Vitiuk}\ \emph {et~al.}(2020)\citenamefont {Vitiuk},
  \citenamefont {Bravina},\ and\ \citenamefont {Zabrodin}}]{Vitiuk:2019rfv}%
  \BibitemOpen
  \bibfield  {author} {\bibinfo {author} {\bibfnamefont {O.}~\bibnamefont
  {Vitiuk}}, \bibinfo {author} {\bibfnamefont {L.~V.}\ \bibnamefont
  {Bravina}},\ and\ \bibinfo {author} {\bibfnamefont {E.~E.}\ \bibnamefont
  {Zabrodin}},\ }\href {https://doi.org/10.1016/j.physletb.2020.135298}
  {\bibfield  {journal} {\bibinfo  {journal} {Phys. Lett. B}\ }\textbf
  {\bibinfo {volume} {803}},\ \bibinfo {pages} {135298} (\bibinfo {year}
  {2020})},\ \Eprint {https://arxiv.org/abs/1910.06292} {arXiv:1910.06292
  [hep-ph]} \BibitemShut {NoStop}%
\bibitem [{\citenamefont {Becattini}\ and\ \citenamefont
  {Lisa}(2020)}]{Becattini:2020ngo}%
  \BibitemOpen
  \bibfield  {author} {\bibinfo {author} {\bibfnamefont {F.}~\bibnamefont
  {Becattini}}\ and\ \bibinfo {author} {\bibfnamefont {M.~A.}\ \bibnamefont
  {Lisa}},\ }\href {https://doi.org/10.1146/annurev-nucl-021920-095245}
  {\bibfield  {journal} {\bibinfo  {journal} {Ann. Rev. Nucl. Part. Sci.}\
  }\textbf {\bibinfo {volume} {70}},\ \bibinfo {pages} {395} (\bibinfo {year}
  {2020})},\ \Eprint {https://arxiv.org/abs/2003.03640} {arXiv:2003.03640
  [nucl-ex]} \BibitemShut {NoStop}%
\bibitem [{\citenamefont {Karpenko}()}]{Karpenko:2021wdm}%
  \BibitemOpen
  \bibfield  {author} {\bibinfo {author} {\bibfnamefont {I.}~\bibnamefont
  {Karpenko}},\ }\Eprint {https://arxiv.org/abs/2101.04963} {arXiv:2101.04963
  [nucl-th]} \BibitemShut {NoStop}%
\bibitem [{\citenamefont {Huang}\ \emph {et~al.}()\citenamefont {Huang},
  \citenamefont {Liao}, \citenamefont {Wang},\ and\ \citenamefont
  {Xia}}]{Huang:2020dtn}%
  \BibitemOpen
  \bibfield  {author} {\bibinfo {author} {\bibfnamefont {X.-G.}\ \bibnamefont
  {Huang}}, \bibinfo {author} {\bibfnamefont {J.}~\bibnamefont {Liao}},
  \bibinfo {author} {\bibfnamefont {Q.}~\bibnamefont {Wang}},\ and\ \bibinfo
  {author} {\bibfnamefont {X.-L.}\ \bibnamefont {Xia}},\ }\href@noop {}
  {}\Eprint {https://arxiv.org/abs/2010.08937} {arXiv:2010.08937 [nucl-th]}
  \BibitemShut {NoStop}%
\bibitem [{\citenamefont {Adamczyk}\ \emph {et~al.}(2017)\citenamefont
  {Adamczyk} \emph {et~al.}}]{STAR:2017ckg}%
  \BibitemOpen
  \bibfield  {author} {\bibinfo {author} {\bibfnamefont {L.}~\bibnamefont
  {Adamczyk}} \emph {et~al.} (\bibinfo {collaboration} {STAR}),\ }\href
  {https://doi.org/10.1038/nature23004} {\bibfield  {journal} {\bibinfo
  {journal} {Nature}\ }\textbf {\bibinfo {volume} {548}},\ \bibinfo {pages}
  {62} (\bibinfo {year} {2017})},\ \Eprint {https://arxiv.org/abs/1701.06657}
  {arXiv:1701.06657 [nucl-ex]} \BibitemShut {NoStop}%
\bibitem [{\citenamefont {Adam}\ \emph {et~al.}(2018)\citenamefont {Adam} \emph
  {et~al.}}]{Adam:2018ivw}%
  \BibitemOpen
  \bibfield  {author} {\bibinfo {author} {\bibfnamefont {J.}~\bibnamefont
  {Adam}} \emph {et~al.} (\bibinfo {collaboration} {STAR}),\ }\href
  {https://doi.org/10.1103/PhysRevC.98.014910} {\bibfield  {journal} {\bibinfo
  {journal} {Phys. Rev. C}\ }\textbf {\bibinfo {volume} {98}},\ \bibinfo
  {pages} {014910} (\bibinfo {year} {2018})},\ \Eprint
  {https://arxiv.org/abs/1805.04400} {arXiv:1805.04400 [nucl-ex]} \BibitemShut
  {NoStop}%
\bibitem [{\citenamefont {Adam}\ \emph {et~al.}(2019)\citenamefont {Adam} \emph
  {et~al.}}]{Adam:2019srw}%
  \BibitemOpen
  \bibfield  {author} {\bibinfo {author} {\bibfnamefont {J.}~\bibnamefont
  {Adam}} \emph {et~al.} (\bibinfo {collaboration} {STAR}),\ }\href
  {https://doi.org/10.1103/PhysRevLett.123.132301} {\bibfield  {journal}
  {\bibinfo  {journal} {Phys. Rev. Lett.}\ }\textbf {\bibinfo {volume} {123}},\
  \bibinfo {pages} {132301} (\bibinfo {year} {2019})},\ \Eprint
  {https://arxiv.org/abs/1905.11917} {arXiv:1905.11917 [nucl-ex]} \BibitemShut
  {NoStop}%
\bibitem [{\citenamefont {Adams}(2021)}]{Adams:2021idn}%
  \BibitemOpen
  \bibfield  {author} {\bibinfo {author} {\bibfnamefont {J.~R.}\ \bibnamefont
  {Adams}} (\bibinfo {collaboration} {STAR}),\ }\href
  {https://doi.org/10.1016/j.nuclphysa.2020.121864} {\bibfield  {journal}
  {\bibinfo  {journal} {Nucl. Phys. A}\ }\textbf {\bibinfo {volume} {1005}},\
  \bibinfo {pages} {121864} (\bibinfo {year} {2021})}\BibitemShut {NoStop}%
\bibitem [{\citenamefont {Wu}\ \emph {et~al.}(2019)\citenamefont {Wu},
  \citenamefont {Pang}, \citenamefont {Huang},\ and\ \citenamefont
  {Wang}}]{Wu:2019eyi}%
  \BibitemOpen
  \bibfield  {author} {\bibinfo {author} {\bibfnamefont {H.-Z.}\ \bibnamefont
  {Wu}}, \bibinfo {author} {\bibfnamefont {L.-G.}\ \bibnamefont {Pang}},
  \bibinfo {author} {\bibfnamefont {X.-G.}\ \bibnamefont {Huang}},\ and\
  \bibinfo {author} {\bibfnamefont {Q.}~\bibnamefont {Wang}},\ }\href
  {https://doi.org/10.1103/PhysRevResearch.1.033058} {\bibfield  {journal}
  {\bibinfo  {journal} {Phys. Rev. Research.}\ }\textbf {\bibinfo {volume}
  {1}},\ \bibinfo {pages} {033058} (\bibinfo {year} {2019})},\ \Eprint
  {https://arxiv.org/abs/1906.09385} {arXiv:1906.09385 [nucl-th]} \BibitemShut
  {NoStop}%
\bibitem [{\citenamefont {Becattini}\ and\ \citenamefont
  {Karpenko}(2018)}]{Becattini:2017gcx}%
  \BibitemOpen
  \bibfield  {author} {\bibinfo {author} {\bibfnamefont {F.}~\bibnamefont
  {Becattini}}\ and\ \bibinfo {author} {\bibfnamefont {I.}~\bibnamefont
  {Karpenko}},\ }\href {https://doi.org/10.1103/PhysRevLett.120.012302}
  {\bibfield  {journal} {\bibinfo  {journal} {Phys. Rev. Lett.}\ }\textbf
  {\bibinfo {volume} {120}},\ \bibinfo {pages} {012302} (\bibinfo {year}
  {2018})},\ \Eprint {https://arxiv.org/abs/1707.07984} {arXiv:1707.07984
  [nucl-th]} \BibitemShut {NoStop}%
\bibitem [{\citenamefont {Acharya}\ \emph {et~al.}(2020)\citenamefont {Acharya}
  \emph {et~al.}}]{Acharya:2019ryw}%
  \BibitemOpen
  \bibfield  {author} {\bibinfo {author} {\bibfnamefont {S.}~\bibnamefont
  {Acharya}} \emph {et~al.} (\bibinfo {collaboration} {ALICE}),\ }\href
  {https://doi.org/10.1103/PhysRevC.101.044611} {\bibfield  {journal} {\bibinfo
   {journal} {Phys. Rev. C}\ }\textbf {\bibinfo {volume} {101}},\ \bibinfo
  {pages} {044611} (\bibinfo {year} {2020})},\ \Eprint
  {https://arxiv.org/abs/1909.01281} {arXiv:1909.01281 [nucl-ex]} \BibitemShut
  {NoStop}%
\bibitem [{\citenamefont {Adam}\ \emph {et~al.}(2021)\citenamefont {Adam} \emph
  {et~al.}}]{Adam:2020pti}%
  \BibitemOpen
  \bibfield  {author} {\bibinfo {author} {\bibfnamefont {J.}~\bibnamefont
  {Adam}} \emph {et~al.} (\bibinfo {collaboration} {STAR}),\ }\href
  {https://doi.org/10.1103/PhysRevLett.126.162301} {\bibfield  {journal}
  {\bibinfo  {journal} {Phys. Rev. Lett.}\ }\textbf {\bibinfo {volume} {126}},\
  \bibinfo {pages} {162301} (\bibinfo {year} {2021})},\ \Eprint
  {https://arxiv.org/abs/2012.13601} {arXiv:2012.13601 [nucl-ex]} \BibitemShut
  {NoStop}%
\bibitem [{\citenamefont {Kornas}(2020)}]{Kornas:2020qzi}%
  \BibitemOpen
  \bibfield  {author} {\bibinfo {author} {\bibfnamefont {F.~J.}\ \bibnamefont
  {Kornas}} (\bibinfo {collaboration} {HADES}),\ }\href
  {https://doi.org/10.1007/978-3-030-53448-6_68} {\bibfield  {journal}
  {\bibinfo  {journal} {Springer Proc. Phys.}\ }\textbf {\bibinfo {volume}
  {250}},\ \bibinfo {pages} {435} (\bibinfo {year} {2020})}\BibitemShut
  {NoStop}%
\bibitem [{\citenamefont {Deng}\ \emph {et~al.}(2020)\citenamefont {Deng},
  \citenamefont {Huang}, \citenamefont {Ma},\ and\ \citenamefont
  {Zhang}}]{Deng:2020ygd}%
  \BibitemOpen
  \bibfield  {author} {\bibinfo {author} {\bibfnamefont {X.-G.}\ \bibnamefont
  {Deng}}, \bibinfo {author} {\bibfnamefont {X.-G.}\ \bibnamefont {Huang}},
  \bibinfo {author} {\bibfnamefont {Y.-G.}\ \bibnamefont {Ma}},\ and\ \bibinfo
  {author} {\bibfnamefont {S.}~\bibnamefont {Zhang}},\ }\href
  {https://doi.org/10.1103/PhysRevC.101.064908} {\bibfield  {journal} {\bibinfo
   {journal} {Phys. Rev. C}\ }\textbf {\bibinfo {volume} {101}},\ \bibinfo
  {pages} {064908} (\bibinfo {year} {2020})},\ \Eprint
  {https://arxiv.org/abs/2001.01371} {arXiv:2001.01371 [nucl-th]} \BibitemShut
  {NoStop}%
\bibitem [{\citenamefont {Ivanov}(2021)}]{Ivanov:2020udj}%
  \BibitemOpen
  \bibfield  {author} {\bibinfo {author} {\bibfnamefont {Y.~B.}\ \bibnamefont
  {Ivanov}},\ }\href {https://doi.org/10.1103/PhysRevC.103.L031903} {\bibfield
  {journal} {\bibinfo  {journal} {Phys. Rev. C}\ }\textbf {\bibinfo {volume}
  {103}},\ \bibinfo {pages} {L031903} (\bibinfo {year} {2021})},\ \Eprint
  {https://arxiv.org/abs/2012.07597} {arXiv:2012.07597 [nucl-th]} \BibitemShut
  {NoStop}%
\bibitem [{\citenamefont {Guo}\ \emph {et~al.}()\citenamefont {Guo},
  \citenamefont {Liao}, \citenamefont {Wang}, \citenamefont {Xing},\ and\
  \citenamefont {Zhang}}]{Guo:2021uqc}%
  \BibitemOpen
  \bibfield  {author} {\bibinfo {author} {\bibfnamefont {Y.}~\bibnamefont
  {Guo}}, \bibinfo {author} {\bibfnamefont {J.}~\bibnamefont {Liao}}, \bibinfo
  {author} {\bibfnamefont {E.}~\bibnamefont {Wang}}, \bibinfo {author}
  {\bibfnamefont {H.}~\bibnamefont {Xing}},\ and\ \bibinfo {author}
  {\bibfnamefont {H.}~\bibnamefont {Zhang}},\ }\href@noop {} {}\Eprint
  {https://arxiv.org/abs/2105.13481} {arXiv:2105.13481 [nucl-th]} \BibitemShut
  {NoStop}%
\bibitem [{\citenamefont {Abdallah}\ \emph {et~al.}()\citenamefont {Abdallah}
  \emph {et~al.}}]{STAR:2021beb}%
  \BibitemOpen
  \bibfield  {author} {\bibinfo {author} {\bibfnamefont {M.~S.}\ \bibnamefont
  {Abdallah}} \emph {et~al.} (\bibinfo {collaboration} {STAR}),\ }\href@noop {}
  {}\Eprint {https://arxiv.org/abs/2108.00044} {arXiv:2108.00044 [nucl-ex]}
  \BibitemShut {NoStop}%
\bibitem [{\citenamefont {Okubo}(2021)}]{Okubo:2021dbt}%
  \BibitemOpen
  \bibfield  {author} {\bibinfo {author} {\bibfnamefont {K.}~\bibnamefont
  {Okubo}} (\bibinfo {collaboration} {STAR}),\ }in\ \href@noop {} {\emph
  {\bibinfo {booktitle} {{19th International Conference on Strangeness in Quark
  Matter}}}}\ (\bibinfo {year} {2021})\ \Eprint
  {https://arxiv.org/abs/2108.10012} {arXiv:2108.10012 [nucl-ex]} \BibitemShut
  {NoStop}%
\bibitem [{\citenamefont {Sa}\ \emph {et~al.}(2012)\citenamefont {Sa},
  \citenamefont {Zhou}, \citenamefont {Yan}, \citenamefont {Li}, \citenamefont
  {Feng}, \citenamefont {Dong},\ and\ \citenamefont {Cai}}]{Sa:2011ye}%
  \BibitemOpen
  \bibfield  {author} {\bibinfo {author} {\bibfnamefont {B.-H.}\ \bibnamefont
  {Sa}}, \bibinfo {author} {\bibfnamefont {D.-M.}\ \bibnamefont {Zhou}},
  \bibinfo {author} {\bibfnamefont {Y.-L.}\ \bibnamefont {Yan}}, \bibinfo
  {author} {\bibfnamefont {X.-M.}\ \bibnamefont {Li}}, \bibinfo {author}
  {\bibfnamefont {S.-Q.}\ \bibnamefont {Feng}}, \bibinfo {author}
  {\bibfnamefont {B.-G.}\ \bibnamefont {Dong}},\ and\ \bibinfo {author}
  {\bibfnamefont {X.}~\bibnamefont {Cai}},\ }\href
  {https://doi.org/10.1016/j.cpc.2011.08.021} {\bibfield  {journal} {\bibinfo
  {journal} {Comput. Phys. Commun.}\ }\textbf {\bibinfo {volume} {183}},\
  \bibinfo {pages} {333} (\bibinfo {year} {2012})},\ \Eprint
  {https://arxiv.org/abs/1104.1238} {arXiv:1104.1238 [nucl-th]} \BibitemShut
  {NoStop}%
\bibitem [{\citenamefont {{F. Becattini, M. Buzzegoli, and A.
  Palermo}}()}]{Becattini:2021suc}%
  \BibitemOpen
  \bibfield  {author} {\bibinfo {author} {\bibnamefont {{F. Becattini, M.
  Buzzegoli, and A. Palermo}}},\ }\href@noop {} {}\Eprint
  {https://arxiv.org/abs/2103.10917} {arXiv:2103.10917 [nucl-th]} \BibitemShut
  {NoStop}%
\bibitem [{\citenamefont {Liu}\ and\ \citenamefont {Yin}()}]{Liu:2021uhn}%
  \BibitemOpen
  \bibfield  {author} {\bibinfo {author} {\bibfnamefont {S.~Y.~F.}\
  \bibnamefont {Liu}}\ and\ \bibinfo {author} {\bibfnamefont {Y.}~\bibnamefont
  {Yin}},\ }\href@noop {} {}\Eprint {https://arxiv.org/abs/2103.09200}
  {arXiv:2103.09200 [hep-ph]} \BibitemShut {NoStop}%
\bibitem [{\citenamefont {Sjostrand}\ \emph {et~al.}(2006)\citenamefont
  {Sjostrand}, \citenamefont {Mrenna},\ and\ \citenamefont
  {Skands}}]{Sjostrand:2006za}%
  \BibitemOpen
  \bibfield  {author} {\bibinfo {author} {\bibfnamefont {T.}~\bibnamefont
  {Sjostrand}}, \bibinfo {author} {\bibfnamefont {S.}~\bibnamefont {Mrenna}},\
  and\ \bibinfo {author} {\bibfnamefont {P.~Z.}\ \bibnamefont {Skands}},\
  }\href {https://doi.org/10.1088/1126-6708/2006/05/026} {\bibfield  {journal}
  {\bibinfo  {journal} {JHEP}\ }\textbf {\bibinfo {volume} {05}},\ \bibinfo
  {pages} {026}},\ \Eprint {https://arxiv.org/abs/hep-ph/0603175}
  {arXiv:hep-ph/0603175} \BibitemShut {NoStop}%
\bibitem [{\citenamefont {Teryaev}\ and\ \citenamefont
  {Usubov}(2015)}]{Teryaev:2015gxa}%
  \BibitemOpen
  \bibfield  {author} {\bibinfo {author} {\bibfnamefont {O.}~\bibnamefont
  {Teryaev}}\ and\ \bibinfo {author} {\bibfnamefont {R.}~\bibnamefont
  {Usubov}},\ }\href {https://doi.org/10.1103/PhysRevC.92.014906} {\bibfield
  {journal} {\bibinfo  {journal} {Phys. Rev. C}\ }\textbf {\bibinfo {volume}
  {92}},\ \bibinfo {pages} {014906} (\bibinfo {year} {2015})}\BibitemShut
  {NoStop}%
\bibitem [{\citenamefont {Deng}\ and\ \citenamefont
  {Huang}(2016)}]{Deng:2016gyh}%
  \BibitemOpen
  \bibfield  {author} {\bibinfo {author} {\bibfnamefont {W.-T.}\ \bibnamefont
  {Deng}}\ and\ \bibinfo {author} {\bibfnamefont {X.-G.}\ \bibnamefont
  {Huang}},\ }\href {https://doi.org/10.1103/PhysRevC.93.064907} {\bibfield
  {journal} {\bibinfo  {journal} {Phys. Rev. C}\ }\textbf {\bibinfo {volume}
  {93}},\ \bibinfo {pages} {064907} (\bibinfo {year} {2016})},\ \Eprint
  {https://arxiv.org/abs/1603.06117} {arXiv:1603.06117 [nucl-th]} \BibitemShut
  {NoStop}%
\bibitem [{\citenamefont {Pang}\ \emph {et~al.}(2012)\citenamefont {Pang},
  \citenamefont {Wang},\ and\ \citenamefont {Wang}}]{Pang:2012he}%
  \BibitemOpen
  \bibfield  {author} {\bibinfo {author} {\bibfnamefont {L.}~\bibnamefont
  {Pang}}, \bibinfo {author} {\bibfnamefont {Q.}~\bibnamefont {Wang}},\ and\
  \bibinfo {author} {\bibfnamefont {X.-N.}\ \bibnamefont {Wang}},\ }\href
  {https://doi.org/10.1103/PhysRevC.86.024911} {\bibfield  {journal} {\bibinfo
  {journal} {Phys. Rev. C}\ }\textbf {\bibinfo {volume} {86}},\ \bibinfo
  {pages} {024911} (\bibinfo {year} {2012})},\ \Eprint
  {https://arxiv.org/abs/1205.5019} {arXiv:1205.5019 [nucl-th]} \BibitemShut
  {NoStop}%
\bibitem [{\citenamefont {Hirano}\ \emph {et~al.}(2013)\citenamefont {Hirano},
  \citenamefont {Huovinen}, \citenamefont {Murase},\ and\ \citenamefont
  {Nara}}]{Hirano:2012kj}%
  \BibitemOpen
  \bibfield  {author} {\bibinfo {author} {\bibfnamefont {T.}~\bibnamefont
  {Hirano}}, \bibinfo {author} {\bibfnamefont {P.}~\bibnamefont {Huovinen}},
  \bibinfo {author} {\bibfnamefont {K.}~\bibnamefont {Murase}},\ and\ \bibinfo
  {author} {\bibfnamefont {Y.}~\bibnamefont {Nara}},\ }\href
  {https://doi.org/10.1016/j.ppnp.2013.02.002} {\bibfield  {journal} {\bibinfo
  {journal} {Prog. Part. Nucl. Phys.}\ }\textbf {\bibinfo {volume} {70}},\
  \bibinfo {pages} {108} (\bibinfo {year} {2013})},\ \Eprint
  {https://arxiv.org/abs/1204.5814} {arXiv:1204.5814 [nucl-th]} \BibitemShut
  {NoStop}%
\bibitem [{\citenamefont {Oliinychenko}\ and\ \citenamefont
  {Petersen}(2016)}]{Oliinychenko:2015lva}%
  \BibitemOpen
  \bibfield  {author} {\bibinfo {author} {\bibfnamefont {D.}~\bibnamefont
  {Oliinychenko}}\ and\ \bibinfo {author} {\bibfnamefont {H.}~\bibnamefont
  {Petersen}},\ }\href {https://doi.org/10.1103/PhysRevC.93.034905} {\bibfield
  {journal} {\bibinfo  {journal} {Phys. Rev. C}\ }\textbf {\bibinfo {volume}
  {93}},\ \bibinfo {pages} {034905} (\bibinfo {year} {2016})},\ \Eprint
  {https://arxiv.org/abs/1508.04378} {arXiv:1508.04378 [nucl-th]} \BibitemShut
  {NoStop}%
\bibitem [{\citenamefont {Lin}(2014)}]{Lin:2014tya}%
  \BibitemOpen
  \bibfield  {author} {\bibinfo {author} {\bibfnamefont {Z.-W.}\ \bibnamefont
  {Lin}},\ }\href {https://doi.org/10.1103/PhysRevC.90.014904} {\bibfield
  {journal} {\bibinfo  {journal} {Phys. Rev. C}\ }\textbf {\bibinfo {volume}
  {90}},\ \bibinfo {pages} {014904} (\bibinfo {year} {2014})},\ \Eprint
  {https://arxiv.org/abs/1403.6321} {arXiv:1403.6321 [nucl-th]} \BibitemShut
  {NoStop}%
\end{thebibliography}

\end{document}